\documentclass [12pt]{article}
\usepackage {amssymb}
\usepackage {amsmath}
\usepackage{graphicx}
\usepackage{cite}
\usepackage {longtable}

\title{The world of strategies with memory}

\author{$^{1}$\textbf{V.M. Kuklin}, $^1$\textbf{V.V. Porichansky}, $^1$\textbf{A.V. Priymak}, $^{1,2}$\textbf{V.V.Yanovsky}}

\begin{document}

 \maketitle

$^{1}$Kharkov National University named after V.N. Karazin, pl. Freedom, 4, 61000, Kharkov, Ukraine

$^{2}$\textit{Institute for single crystals, National Academy of
Science of Ukraine, Nauki Ave 60, 61001 Kharkiv, Ukraine}

\abstract{As part of a generalized "prisoners' dilemma", is considered that the evolution of a population with a full set of behavioral strategies limited only by the depth of memory. Each subsequent generation of the population successively loses the most disadvantageous strategies of behavior of the previous generation. It is shown that an increase in memory in a population is evolutionarily beneficial. The winners of evolutionary selection invariably refer to agents with maximum memory. The concept of strategy complexity is introduced. It is shown that strategies that win in natural selection have maximum or near maximum complexity. Despite the fact that at a separate stage of evolution, according to the payout matrix, the individual gain, while refusing to cooperate, exceeded the gain obtained while cooperating. The winning strategies always belonged to the so-called respectable strategies that are clearly prone to cooperation.}

\section{Introduction}

Understanding the nature of the emergence of cooperative behavior in different systems has been of interest to researchers for several decades. The evolutionary games theory \cite{1s,2s,3s} provides flexible foundations and effective methods for exploring the emergence of collaboration. Among the many game models that are used to explain cooperative behavior, a special place is played by games, which can be considered as a generalization of the prisoners' dilemmas \cite{4s,5s,6s}. The choice of a payout matrix in this case is determined by a simple physical consideration. Cooperation always requires additional resources in comparison with the rejection of cooperation. The tendency to save resources or efforts is manifested in the payout matrix in the fact that each individual interaction, the individual gain, while refusing to cooperate, exceeds the gain while agreeing to cooperate. At each stage of the evolutionary process or generation, the population refuses to use the least successful strategies of the previous generation.
These games serve as a paradigm that led to the discovery of the mechanism of cooperative behavior in theory and experimental observations \cite{7s}.
Starting with the work of Novak and May \cite{8s}, evolutionary games have been widely studied in structured populations, including on regular lattices \cite{9s,10s,11s,12s,13s,14s,15s,16s,17s} and complex networks \cite{18s,19s,20s,21s,22s,23s,24s,25s,26s,27s,28s,29s,30s,31s,32s,33s,34s}.

At present, a number of general, but specific mechanisms have been discovered that lead to cooperation in a wide variety of systems (see, for example, \cite{35s}). Among such mechanisms, it should be noted: voluntary participation \cite{36s}, punishment\cite{37s}, similarity \cite{38s}, heterogeneous activity \cite{39s}, social diversity \cite{40s,41s}, dynamic connections \cite{42s}, asymmetric interaction and the graph of permutations \cite{43s}, migration \cite{44s,45s,46s}, group favoritism \cite{47s}, interdependent relationships \cite{48s}. Using this approach, one can determine the appearance of many diverse properties in evolving populations. Following evolutionary populations, following Darwin, we will understand many objects that obey the following principles. These are 1) the principle of heredity, 2) the principle of variability and 3) natural selection.

In this paper, we analyze the effect of memory on the evolutionary process. We will use the strategy memory in a simplified version compared to that proposed in \cite{49s}. If the action of an object depends not only on the observed situation, but also on previous events, then we assume that the object has memory. In this sense, most biological objects have memory. The main issue that we will discuss in this work is how beneficial it is for a population to increase memory in the process of evolution and what consequences this leads to. The depth of memory of population objects by the number of enemy moves that the strategy takes into account when making a decision. The paper considers the Cauchy problem of the evolution of all strategies with a memory depth not exceeding a certain number. In other words, competition in the original population of all possible strategies with limited memory above.

The second issue that is addressed in the work is related to the discussion of the properties of competing strategies leading to a change in the dominant strategies of the population in the process of evolution. As a characteristic of strategies, their complexity is introduced and used. The main question is: is the complexity of strategies evolutionarily beneficial? At an intuitive level, the answers to these questions seem obvious. In modeling the interaction of strategies, a single-particle approximation was used, in which all agents of the population professing one of the possible strategies were combined into a single cluster. The interaction was between clusters or strategies. In other words, it is precisely the strategies that interact. In addition, each strategy interacts with each, including itself. Three types of populations are considered. Populations without memory, populations with a memory depth of 1 and 2. Over-exponential growth in the number of strategies with increasing memory depth greatly limits the ability to model populations with more memory.

In each case, in the initial population there are all strategies with memory that do not exceed the specified. So, for example, at a depth of memory of 2, all strategies with a memory of 2, 1, and 0 are present. As a result of numerical simulation, it is shown that an increase in memory in a population is evolutionarily beneficial. The winners of evolutionary selection invariably refer to agents with maximum memory. Strategies that win natural selection have maximum or near maximum complexity. Along the way, it was found that in such populations, the winning strategies belonged to "respectable" strategies, prone to cooperation. In a certain sense, we can say that cooperative behavior in such cases is automatically established. It can be expected that this is the universal trend. In populations with limited memory above, the competition of all possible strategies in the initial population leads to the domination of respectable strategies during the evolution. A further increase in the depth of memory leads to a new problem when the number of agents in the population is less than the number of possible strategies. A consequence of this is also discussed in the conclusion of this work.

\section{Strategies with memory}

All strategies that use memory will be considered. We will introduce the space of such strategies. The strategy is this rule on that motion is determined by the well-known values of motions of the opponent. For their classification, the depth of memory is used. Under the depth of memory is understood the number of previous motions that are used by strategy for implementation of motion. We will begin with the simplest strategies that does not use memory. It means that such strategies carry out motion, being base on only the looked after the motion of the opponent. In this case the possible looked after motion it $0$ or $1$ ( $0$ = refuse, $1$ = co-operation ). Accordingly, for a description of a separate strategy, we need to describe the certain rule of the answer for these motions. In an order to describe all such strategies it is needed to create all rules of reaction on the value of motions of opponent. We will begin with the method of record of some strategy at the depth of memory $0$. Clear, that strategy can be set by a next table
\[\begin{tabular}{|c|c|c|}
  \hline
  \text{Possible opponent move} & 0 & 1 \\ \hline
     & $\Downarrow$ & $\Downarrow$ \\ \hline
  \text{Response strategy } & 0 & 0 \\
  \hline
\end{tabular}
\]

It is easy to understand that if arrange about the order of record of possible values of motions of opponent, for example, in the lexicographic order (as in a table in an overhead line), then for description of rules of action of strategy it is enough to know lower range or sequence of zeros and ones. In the example given above, it is a sequence $00$. Because a specific sequence corresponds to every strategy, then she can be used and as the name of the corresponding strategy. In this case, the name and determines the rule of action of strategy. Thus, strategies with a zero-depth of memory are determined by  $0,1$-sequences from two elements. Then the name of every strategy with a zero-depth of memory is determined by a binary number with two signs.

It is clear that the names of all strategies in the absence of memory are numbers from $00$ to $11$, whence it is evident that there are four such strategies. As an example, we give a strategy with the name $10$, which operates according to the following rules.
\[\begin{tabular}{|c|c|c|}
  \hline
  \text{Possible opponent move} & 0 & 1 \\ \hline
     & $\Downarrow$ & $\Downarrow$ \\ \hline
  \text{Response strategy } & 1 & 0 \\
  \hline
\end{tabular}
\]

There are 4 such sequences $00$, $01$, $10$, $11$, and each corresponds to a certain strategy. Among these strategies, two are banal. It is an extremely aggressive strategy $00$ and thoughtlessly conciliating $11$. It remains only to discuss the choice of the first motion that comes true, not on the rules indicated higher. This choice takes place from two variants. The first motion can be $1$ or $0$. Therefore it comfortably to plug the first motion in the description of the strategy. For this purpose, we will specify it in brackets before the name of the strategy. For example, name $[1]10$ and $[0]10$ means that the first motion, according to $1$ and $0$, and 3 further actions come true in obedience to strategy $10$. It comfortably to examine every strategy with the pointing of the first motion as a separate strategy. Then in default of memory, the number of all strategies is equal to 8.

We will consider now all strategies with memory about one previous motion. Such strategies must take into account two motions of the opponent. Previous and looked after. Then the number of possible variants of motions of opponent increases, and the strategy must determine return motion taking into account the previous motion of the opponent. Again we will dispose of all possible pairs of motions of the opponent in the lexicographic order:
\[\begin{tabular}{|c|c|c|c|c|}
  \hline
  \text{Possible opponent move} & $00$ & $01$ & $10$ & $11$ \\ \hline
     & $\Downarrow$ & $\Downarrow$& $\Downarrow$&$\Downarrow$ \\ \hline
  \text{Response strategy} & . & . & .& . \\
  \hline
\end{tabular}
\]
To describe the strategy, each pair of moves of the opponent needs to match $0$ or $1$.

In other words, replace the points in the table with the characters $0$ or $1$. Again, with a fixed order of recording the possible moves of the opponent, each strategy is determined $0,1$-sequence of 4 elements. The name of the corresponding strategy can be chosen again, coinciding with the rule of its action. So the names of strategies with a memory depth of 1 matches a binary number with 4 digits, or with a sequence of 0 and 1 with 4 elements.

The trivial aggressive strategy is called $0000$. In total, there are as many such strategies again as there are numbers from $0000$ to $1111$. In other words, there are 16. There are similar methods for determining strategies with a memory depth of $k$. Such strategies will be determined by a binary number with $2^{k+1}$ digits.    Thus, the strategy space with the memory depth $k$ comprises $0,1$-sequences of $2^{k+1}$  elements.

However, the above description of strategies with a memory depth of  $k \geq 1$ is not complete.  The reason is that after the first move of the enemy, we know only one thing the observed value and the value of the previous move is missing.  Therefore not enough data for applying the specified policy rules.  So we have to specify the rule of how to make a move when data is incomplete.  To do this, it is natural to use one of the strategies with memory $0$. In other words, you should specify a strategy that does not use memory, which we will use before information about the previous course of the enemy appears. Then the full number of strategies is natural.

The name of the strategy increases and changes.  In the name, we must first indicate the strategy for determining the first move in the absence of data on the previous move (i.e., the name strategies with memory $0$) and then the name of the strategy with memory $1$. Thus, the name (and rules) strategies with the memory of one previous opponent's move look like, for example, as $[01] 0111$.  The first two digits in brackets are the name (and rules) of the strategy with memory $0$, and the next four are the rules of the player's moves with the memory of one the course of the enemy.  For convenience, we will attribute strategies with less memory left and enclose in square brackets.  Considering each such rule as separate strategy, you can easily calculate the number of such strategies.  Consequently, the total number of strategies with the memory of one previous opponent move is equal to $2^2 \times 2^4$. Besides, each strategy can start the game from some first move.  In other words, it can start from $0$ or $1$. It is convenient to consider strategies that make different first moves $0$ or $1$ as different strategies.

Then the number  strategies doubled $2 \times 2^2 \times 2^4 = 128$. The name of such strategies will look,
 for example, like $[0] [01] 0111$; this strategy will start the game from move $0$. In particular, the well-known eye for an eye strategy in these terms corresponds to strategy $[1] [01] 0011$. It should be noted that with knowledge of the depth of memory, in this case, equal to $k = 1$, you may not even use parentheses.  Even if they are absent according to a known record strategies uniquely established rule of action of strategies

All strategies with a memory of two enemy moves, or in the general case with a memory of $k$ enemy moves, are listed in exactly the same way.  Important to emphasize  that the number of strategies $N_k = 2 \times 2^2 \times 2^3 \cdots 2^{2^{k+1}}=2^{(2^{k+2}-1)}$ grows super exponentially  with increasing length or depth of memory $k$.

Let us return to the consideration of strategies with a depth of memory $1$. We note that that among these strategies there are strategies with zero memory depth.  Indeed, if the strategy acts the same in different previous opponent's moves, then she does not actually use information about the previous move or loses the memory of a previous move.  Accordingly, such strategies coincide with strategies with zero memory depth.  The rules for strategies equivalent to zero-memory strategies are defined in the following table.
\[\begin{tabular}{|c|c|c|c|c|}
  \hline
  \text{Possible opponent move} & 00 & 01 & 10 & 11 \\ \hline
     & $\Downarrow$ & $\Downarrow$& $\Downarrow$&$\Downarrow$ \\ \hline
  \text{Response strategy } & $x_1$ & $x_2$ & $x_1$& $x_2$ \\
  \hline
\end{tabular}
\]
Where $x_1$ and $x_2$ take the values $\{0, 1\}$.  It is easy to see from the table that such strategies act independently of the opponent's previous move.  This means that strategies $x_1 x_2 x_1 x_2$ are equivalent to strategies with zero memory $x_1 x_2$.  Thus, among strategies with names of four-digit binary numbers, there are strategies equivalent to all strategies with no memory.  So strategies $0000 \sim 00$, $0101 \sim 01$, $1010 \sim 10$ and $1111 \sim 11$  Therefore, when writing the names of strategies with four-digit binary numbers, all strategies with a memory depth less than or equal to $1$ are present among them. It is easy to understand that this convenient property will be preserved even  Description of strategies with greater depth of memory.  So the names of strategies with a depth of memory $k$ contain all strategies equivalent to strategies with a depth of memory $k-1$, ... $0$. This property should be considered when conducting games between strategies.  Thus, we have identified and listed all the strategies with a certain finite depth of memory.  Therefore, further, when it comes to strategies with a memory depth $k$, we will remember that they include all strategies with a smaller memory depth.  Among these strategies, there are both primitive and complex strategies.  Now we discuss in more detail the concept of complexity of strategies.

\section{Complexity of strategies}

In the previous section, strategies are described by $0,1$ - sequences of a certain length or sequences containing a certain number of members. So, for the memory depth $0$, taking into account the first move, there are 8 such sequences, and for the memory depth $k$, their $N_k = 2^{(2^{k+2}-1)}$. Of course, among these strategies there are strategies equivalent to all strategies with less depth of memory. The number of such strategies is also determined by their depth of memory. For subsequent purposes, we discuss such a property of $0,1$ sequences as complexity. This is an exceptionally deep concept that finds important applications in physics and mathematics. In particular, the concept of randomness is closely related to the concept of complexity 5 (see, for example, \cite{50s}). There are a large number of different options for introducing complexity, some of which can be found in \cite{51s}. The most natural choice of this characteristic can be considered as Kolmogorov complexity \cite{52s}. However, we are encountered with the big problem of calculating Kolmogorov complexity. Therefore, we use a different approach for describing the complexity of finite $0,1$-sequences, which is based on the comparative complexity of functions, in particular polynomials. It is based on the understanding that polynomials of a higher degree are more complicated than of a lower degree.

For a more consistent formulation of complexity, we use, following \cite{54s}, the theory of monads. By a monad we mean a finite set $M$ and a map $A$ of this finite set into itself. In other words, each point of this finite set is associated with another point. A monad is assigned a graph whose vertices have a finite set $M$, and oriented edges connect the vertices in accordance with their mapping $A$. Exactly one edge leaves each vertex $x$ in this graph and it leads to the vertex $A x$.

\begin{figure}
  \centering
  \includegraphics[width=4 cm]{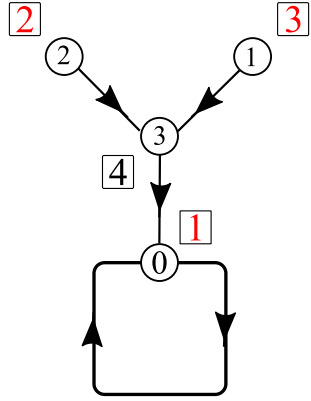}\\
  \caption{Graph of strategies with zero memory. Where, for compactness, the vertices are shown by circles, inside of which the names of strategies are encoded in the decimal system. So, the sequence $00$ is designated as 0, vertex $01$ as 1, vertex $10$ as 2, and finally $11$ as 3. The squares show places occupied by the strategies in the first generation competition.}\label{fg1}
\end{figure}

$0, 1$-sequences $x=x_1 x_2 \ldots x_n$ can also be considered as a function that maps the integer value $i$ to the value $x_i \in \{0,1\}$. As a rule, when recording a sequence, we will not use a comma as a separator. With the selected binary alphabet, this does not lead to confusion. The introduction of complexity of functions dates back to Newton's ideas. To do this, he suggested to use a function differences. In our case, we define such a map by the difference operator $A : M \to M$
\[y=Ax\]
Where the elements of the sequence $y=y_1 y_2 \ldots y_n$ are determined by the differences
\[y_i = x_{i+1}-x_i\]
Where $i=1, 2, \ldots , n$ is the number of the element in the sequence. When calculating the elements of the sequence $y$, we will use the sequence cyclic condition $x$, counting $x_{n+1}=x_1$. In other words, we can speak of periodic sequences of period $n$. Thus, we come to the monad of strategies. Strategy $x$ maps to strategy $y$. Consider graphs of strategies, starting with a shallow depth of memory. The main property of these graphs is that only an edge comes out of each vertex. For sequences of length $n$, the graph contains $2^{n}$ vertices. We start with zero memory depth $k = 0$. Such strategies coincide with sequences of two elements $n=2^{0+1}$. The strategy graph corresponding to this case is shown in Fig.\ref{fg1}.

A characteristic feature of this graph is the presence of a cycle of unit length. The length of the cycle is equal to the number of vertices in the cycle. The notation for this graph is $O_1 \ast T_4$ in accordance with \cite{54s}. Where $O_1$ means a cycle of unit length, and $T_4$ - a binary tree with 4 vertices. It can be proved that the strategy graph will have only one $O_1$ cycle.

Return to the observation of Newton, which consisted in the fact that if the function is constant, then the first differences will be zeros. If the first differences are constant, then the function will be a polynomial of the first degree. And if the second difference is constant, then no more than the second... This observation allows us to formulate complexity as the remoteness of the graph vertices from the root of the tree or cycle \cite{54s}. The farther the vertex is from the root of the tree, the more complicated the strategy. We will use this definition of complexity in this paper.

Accordingly, among the strategies with zero memory, the simplest strategy is the most aggressive strategy $00$ (see Fig.\ref{fg1}). This strategy corresponds to a constant function of argument i, which takes a zero value. A more complex strategy $11$ is a mindlessly compromising strategy. This strategy coincides with the constant function of the argument $i$, which takes the value $1$. "Differentiation" A of this function translates it into a function that takes 0 value. Strategies $01$ and $10$ define linear functions. Indeed, strategy $01$ corresponds to the linear function $x(t)=(t + 1) \mod 2$, and strategy $10$ of the linear function $x(t)=t \mod 2$. It is easy to verify, for example, that the values of the function $x(t)=t+1 \mod 2$ for integers $t$ gives a periodic sequence with period 2 and $x_1 =x(1)=0$, $x_2 =x(2)=1$. It is similarly easy to verify that sequence $10$ corresponds to the values $x(t)=t \mod 2$ at integer points $x_1 =x(1)=1$, $x_2 =x(2)=0$. This coincides with the intuitive conclusion that constant functions are simpler than linear ones. Linear functions are a special case of polynomials of degree less than $p$. As was proved by Newton, if a function satisfies $A^p x =0$, then x is a polynomial of degree less than $p$. We will use this property below to determine polynomials that correspond to more complex strategies.
\begin{figure}
  \centering
  \includegraphics[width=7 cm]{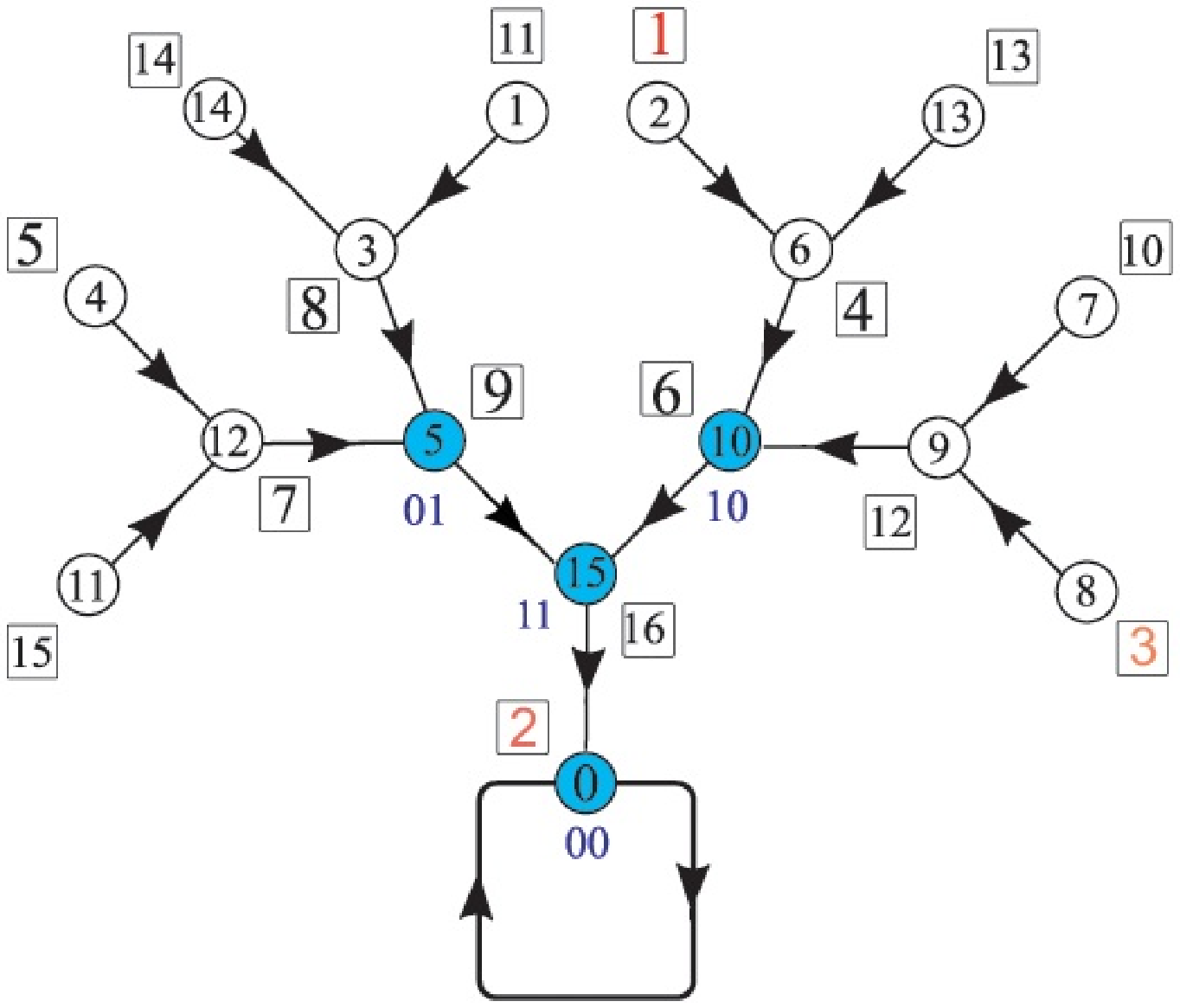}\\
  \caption{Strategies graph with memory depths of $1$ and $0$. Strategies equivalent to strategies 0 of memory depth correspond to vertices marked in blue. The coding of strategies and, accordingly, vertices is carried out as before by writing down the names of strategies in the decimal system of calculus. Vertices have names from $0000$ to $1111$.}\label{fg2}
\end{figure}

Now turn to the monad of strategies with a depth of memory 1. It is a graph shown in Fig.\ref{fg2}. It is easy to see that the structure of the graph corresponds to $O_1 \ast T_{16}$. Let's discuss the location on this graph of strategies equivalent to strategies with zero memory depth. These strategies are shown in Fig.\ref{fg2} with vertices shaded in blue. Next to them in blue are the names of equivalent strategies with zero memory. Thus, these strategies are the simplest in this monad. This coincides exactly with the fact that the complexity of strategies as functions $i$ increases with distance from the root of the graph. The remaining strategies are further removed from the root and, accordingly, more complex. We can verify that the next level of the graph corresponds to polynomials of degree 2 and the upper to polynomials of degree 3. Indeed, acting on the vertex of the 5th level $x^{(5)}$) by the operator $A$ by definition, we pass to the vertex of the 4th level $A x^{(5)}=x^{(4)}$. This is an obvious consequence of the tree structure.

Similarly, the vertex $x^{(4)}$ under the action of $A$ descends to a level below $A x^{(4)}=x^{(3)}$. Repeating the action again, we go to the level below $A x^{(3)}=x^{(2)}$ and, finally, get $A x^{(2)}=x^{(1)} = 0$. Combining these equalities, we get $A^{4} x^{(5)}=0$. We can say that 0 is an attractor of motions induced by the mapping $A$. Then, according to Newton's proof, the vertex of the fifth level $x^{(5)}$ coincides with a polynomial of degree less than $4$.

\begin{figure}
  \centering
  \includegraphics[width=10 cm]{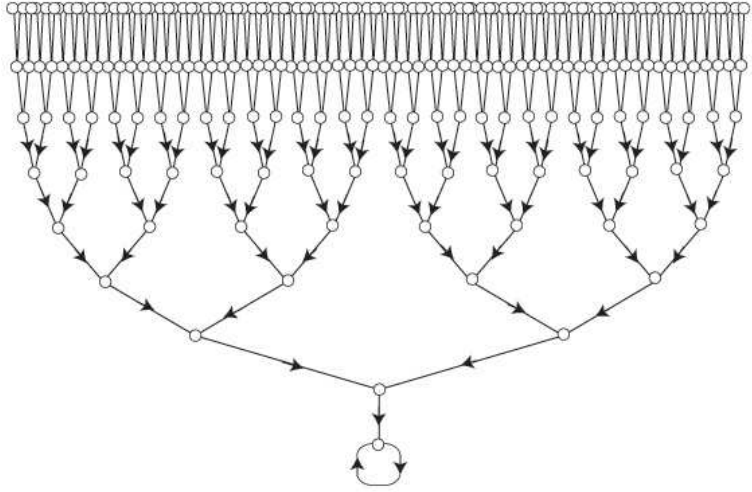}\\
  \caption{Strategy graph with a memory depth of 2. Of course, it is not possible to indicate the names of strategies in such a figure. They are listed in a separate table in the next section.}\label{fg3}
\end{figure}

Now we discuss the monad graph corresponding to strategies with memory depth $k$. In this case, the length of the $0,1$-sequence of the defining strategy is  $n=2^{k+1}$. The total number of such strategies is $N_k = 2^{2^{k+1}}$. It can be proved that for $n=2^{k+1}$ the structure of the strategy graph coincides with $O_1 \ast T_{2^{2^{k+1}}}$. Naturally, at the root there are still strategies equivalent to strategies with zero memory depth, higher with memory depth 1, and so on up to the level corresponding to the last level of the strategy graph with depth $k - 1$, i.e. $O_1 \ast T_{2^{2^{k}}}$. Thus, the number of strategies, $2^{2^{k+1}}-2^{2^{k}}=2^{2^{k}}(2^{2^{k}} -1)$. As they move away from the root, strategies are becoming more complex and correspond to polynomials of an ever higher degree. As an example, in Fig.\ref{fg3} we give the graph corresponding strategies of length $n=2^{3}=8$ ($k=2$), which coincides with $O_1 \ast T_{256}$. Thus, the complexity of strategies can be determined by the value of the level of the graph to which the vertex corresponding to this strategy belongs.

It is important to note that the number of possible strategies with increasing memory increases top exponentially. This leads to deep computational difficulties associated with a lack of resources in the study of the interaction of such strategies. In addition, it is interesting to note that the number of strategies implemented in finite systems is limited more by the number of participants, and not by the number of possible strategies. In other words, in finite systems, a new strategy can always appear and be used in the process of evolution. Life is full of new ideas. This conclusion plays an important role in the numerical simulation of the interaction of a finite number of population objects.

\section{Interaction strategies}

Life in a population and its evolution to a certain extent is determined by the nature of the interaction of strategies of population objects. The simplest case is the pairing of strategies. There are many options for implementing such an interaction. The simplest option is that each strategy interacts with each, including itself. This interaction option can be implemented with a relatively small number of objects. The reason for this is the finiteness of the lifetime of the population object.

Indeed, some characteristic time $\Delta t$ is spent on the interaction of a pair of strategies and, accordingly, time $n^2 \Delta t$ will be spent on the interaction of $n$ strategies with each other. With increasing $n$, the time $n^2 \Delta t$ may exceed the lifetime of the object. Another way of pairwise interaction is when an adversary is chosen randomly among the whole set of strategies, assuming they are equally probable. Another general method that does not use randomness can be implemented using a network of interactions. Interactive strategies will be connected in the graph of this network. It can be generalized, taking into account the interaction of distant vertices with some weight, including probabilistic. You can also there is a spatial structuring of populations \cite{7s}, \cite{9s}, \cite{13s}, \cite{19s}, \cite{38s}, \cite{55s}, in this case, geometric structures of cooperation can arise in space \cite{8s}, \cite{56s}. Another important circumstance that affects the nature of the interaction of strategies has already been noted. This is the finiteness of the many objects that make up the population. In this case, the number of strategies can significantly exceed the number of objects of population. Then the interaction can only occur between part of the strategies. In this work, we will assume that the object in the process of life does not change the strategy and interacts with each strategy of the population, including itself. In other words, we can say that the one-particle approximation of the interaction of strategies is considered -- without taking into account the number of carriers of the strategy.

In order to establish the result of the interaction of strategies, we define a payout matrix. Recall that the dilemma of the two-player prisoner is that each player can choose between cooperation $(1)$ or failure $(0)$. Depending on the opponent's strategy, the selected player receives $a_{11}$ if both cooperate;  $a_{22}$ - if both refuse; $a_{12}$ - if the chosen one cooperates and the opponent refuses; and $a_{21}$ - if the chosen one refuses, the enemy cooperates, where $ a_{21} > a_{11} > a_{22}> a_{12}$ and $2 a_{11} > a_{21} + a_{12}$. In this work, we use the values of the Axelrod payout matrix $M_1$ \cite{57s},
\[\begin{tabular}{|c|c|c|}
  \hline
  & \text{Cooperation} &  \text{Refusal}  \\ \hline
     \text{Cooperation} & 3,3 & 0,5\\ \hline
  \text{Refusal} & 5,0 & 1,1 \\
  \hline
\end{tabular}
\]
Thus, the result of the interaction of strategies will be determined by this matrix. Using the interaction between all strategies with a finite depth of memory, we first establish whether strategies with more memory receive an evolutionary advantage. In addition, it is interesting to study how the complexity of strategies affects the evolutionary advantages of strategies. In other words, is there a reason for the complexity of the systems.

Below we simulate the process of evolution of strategies with memory. For simplicity, the principle of variability will be taken into account in a simple version, assuming that all strategies with a memory depth less than or equal to k are implemented in the population. Since in this case all strategies are taken into account, then other strategies will not appear in the process of evolution. The principle of heredity will consist in transferring winning strategies to descendants. The principle of natural selection is realized by eliminating or destroying losing strategies. Naturally, such a simplified version of evolution can be complicated in many ways. Some of which will be discussed later.

Natural selection is implemented as follows. Let all the strategies interact with each other in a circular system in accordance with an iterated game with a prisoners dilemma. The number of interactions of two strategies in one generation is chosen equal for all equal to n. Actually, the choice of a large number of interactions between the two strategies is designed to exclude the influence of the first move \cite{60s}. As a result of such a competition, strategies gain points in accordance with the above payout matrix. After that, the losing strategy, and possibly several strategies with the minimum number of points, drop out of the next generation. Further, the points of evolutionary advantages are reset and the next round of interactions between the remaining strategies is carried out, corresponding to the formation of new generation strategies.

\section{Collective Variables}

Considering the evolution of strategies, one can control the behavior of each strategy only with a shallow depth of memory. The number of strategies with increasing depth of memory grows superexponentially and individual tracking of strategies becomes not feasible. For example, for memory depth 2, the number of strategies that participate in evolution is 30824. Therefore, you need to enter collective or coarse variables that allow you to monitor certain groups of strategies, united by certain qualities or properties. For us, such properties as the depth of strategy memory and the complexity of strategies will be important. Therefore, we will use the number of strategies $a_i$ with the $i$-th memory depth and the number of strategies $n_i$ with the $i$-th complexity as coarse variables. So for example, $a_1$ is the number of strategies with a memory depth of 1, and $n_3$ is the number of strategies of complexity 3. Such coarse variables allow you to control the change in memory and complexity of strategies of large populations of strategies during evolution.

\section{World without memory}

We begin by discussing the evolution of the simplest world with a memory depth of 0 or a world without memory. Let each strategy interact with another strategy $n = 100$ times in the framework of the iterated prisoners dilemma. The set of points is determined by the payout matrix above, and is added up. Each strategy in one game responds to the first move of the selected opponent, and in another starts, making the first move in the game with the same opponent. In the games that she starts, there are two possibilities to make the first move is to choose 0 or 1. A strategy that makes a certain first move is considered as a separate strategy. After the games are held between all such strategies, including oneself, the strategies are distributed among the occupied places in according to the points earned.
\begin{figure}
  \centering
       \includegraphics[height=4.5 cm]{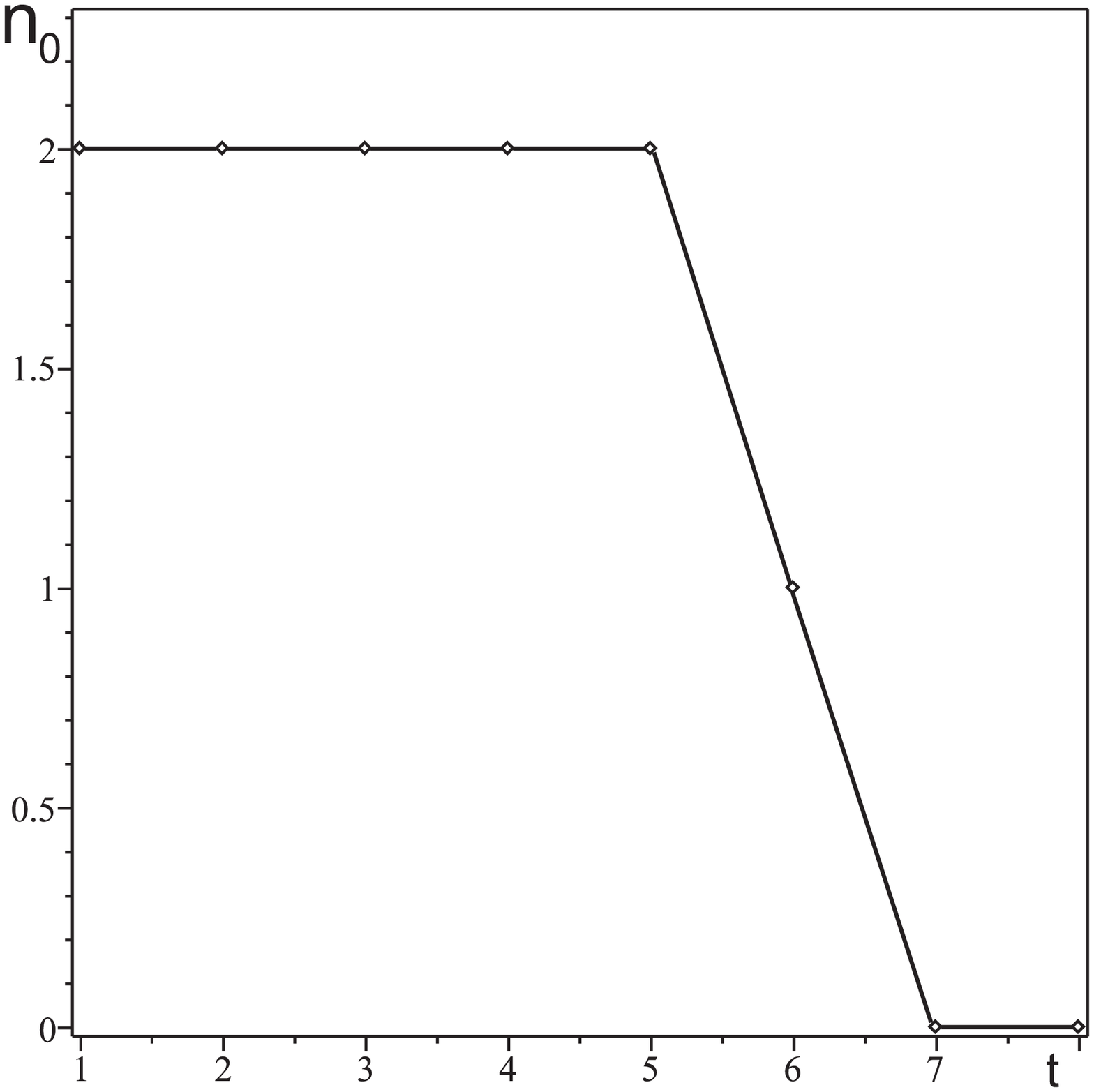}
       \includegraphics[height=4.5 cm]{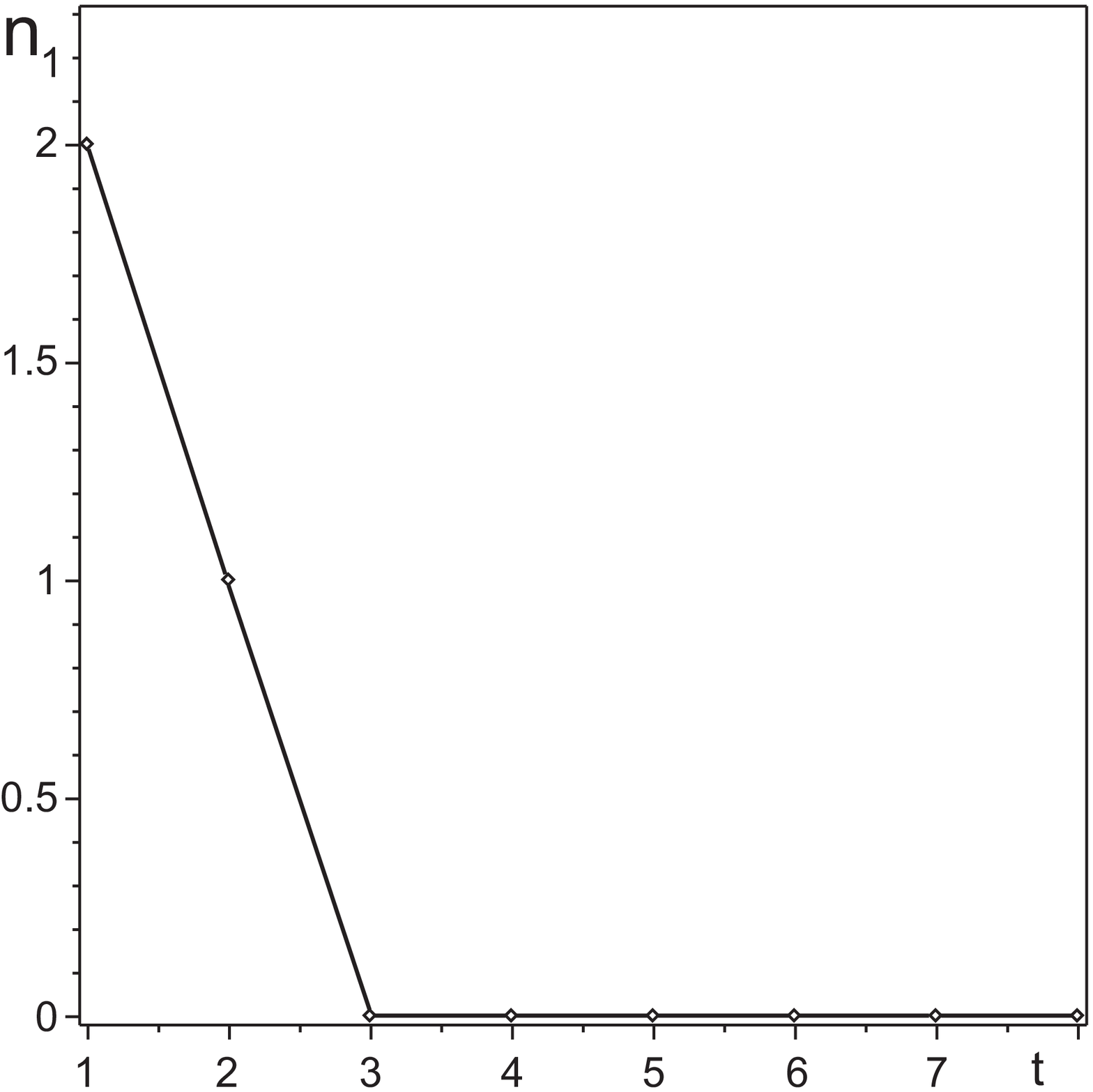}
        \includegraphics[height=4.5 cm]{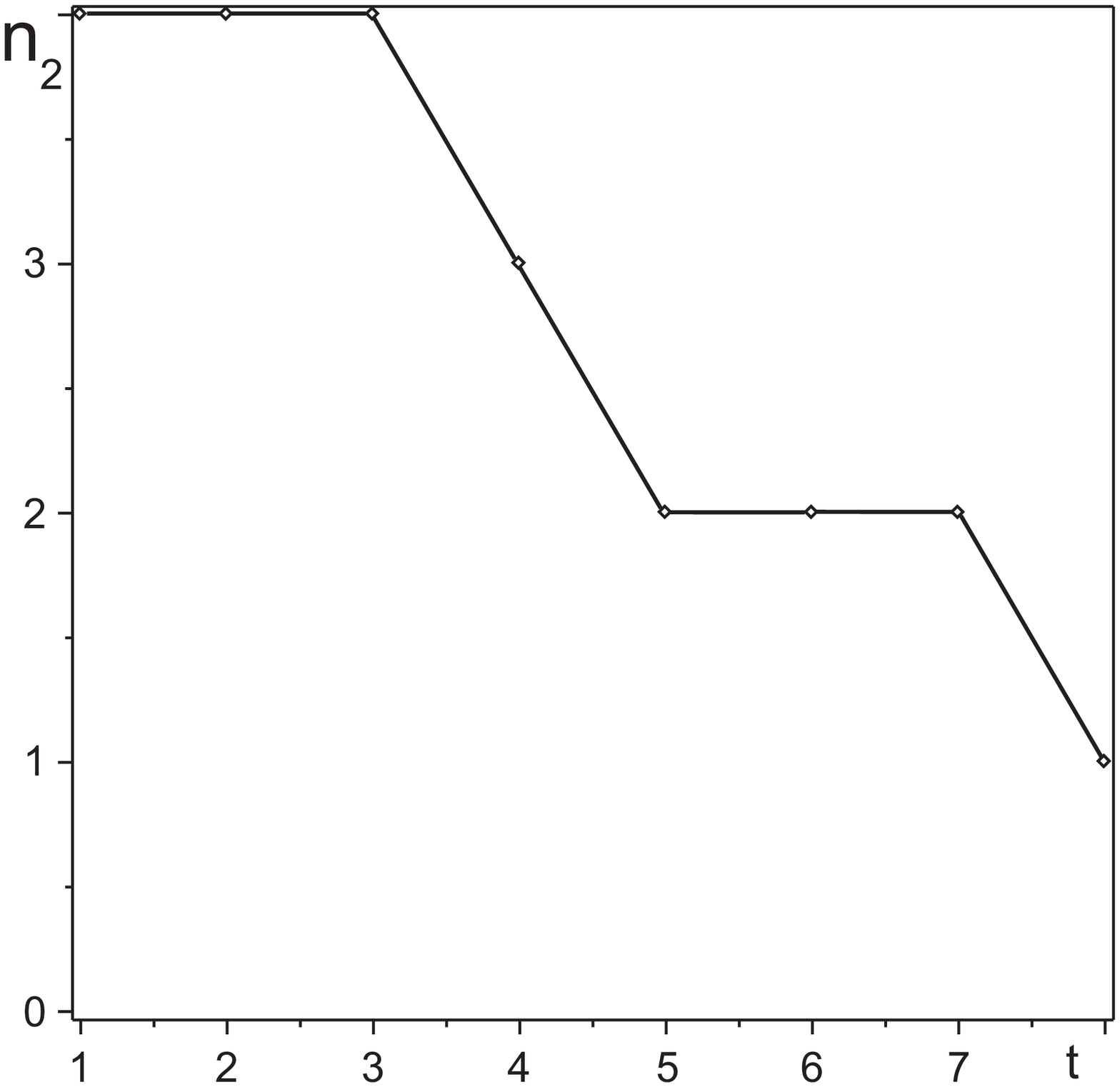}\\
  \caption{Left - change $n_0$ - the number of strategies of zero complexity. In the middle is the evolution of $n_1$- the number of strategies of unit complexity, on the right - $n_2$ - of complexity 2. Note that the points are connected by lines for illustration only and the lines do not play any meaning. Time is discrete.}\label{fg4}
\end{figure}

The first place is occupied by the strategy with the highest total points. The strategy or strategies with the minimum number of points are excluded and are not passed on to the next generation. The remaining strategies are passed on to the next generation and enter the competition again with initial zero points of evolutionary advantages. These strategies can be considered as descendants of the previous generation.

In this simple world, the number of strategies is quite small (4 strategies, and taking into account the first moves are $N_0 = 8$ strategies). Therefore, it is possible to follow all the strategies. However, in it we will use the collective variables discussed above. In this world, all strategies have 0 memory depths and therefore the variables $a_0 (t)$ simply track the number of strategies $a_0 (t) = N_0 (t)$. It is clear that when one losing strategy is removed at each stage of evolution, their number decreases linearly with time $N_0 = (1-t) + 8$. Here $t = 1,2, \ldots , 8$ is the discrete evolution time. All the time of evolution takes 8 stages (or generations), after which one strategy survives and a stationary state sets in.
\begin{figure}
  \centering
  \includegraphics[width=5 cm]{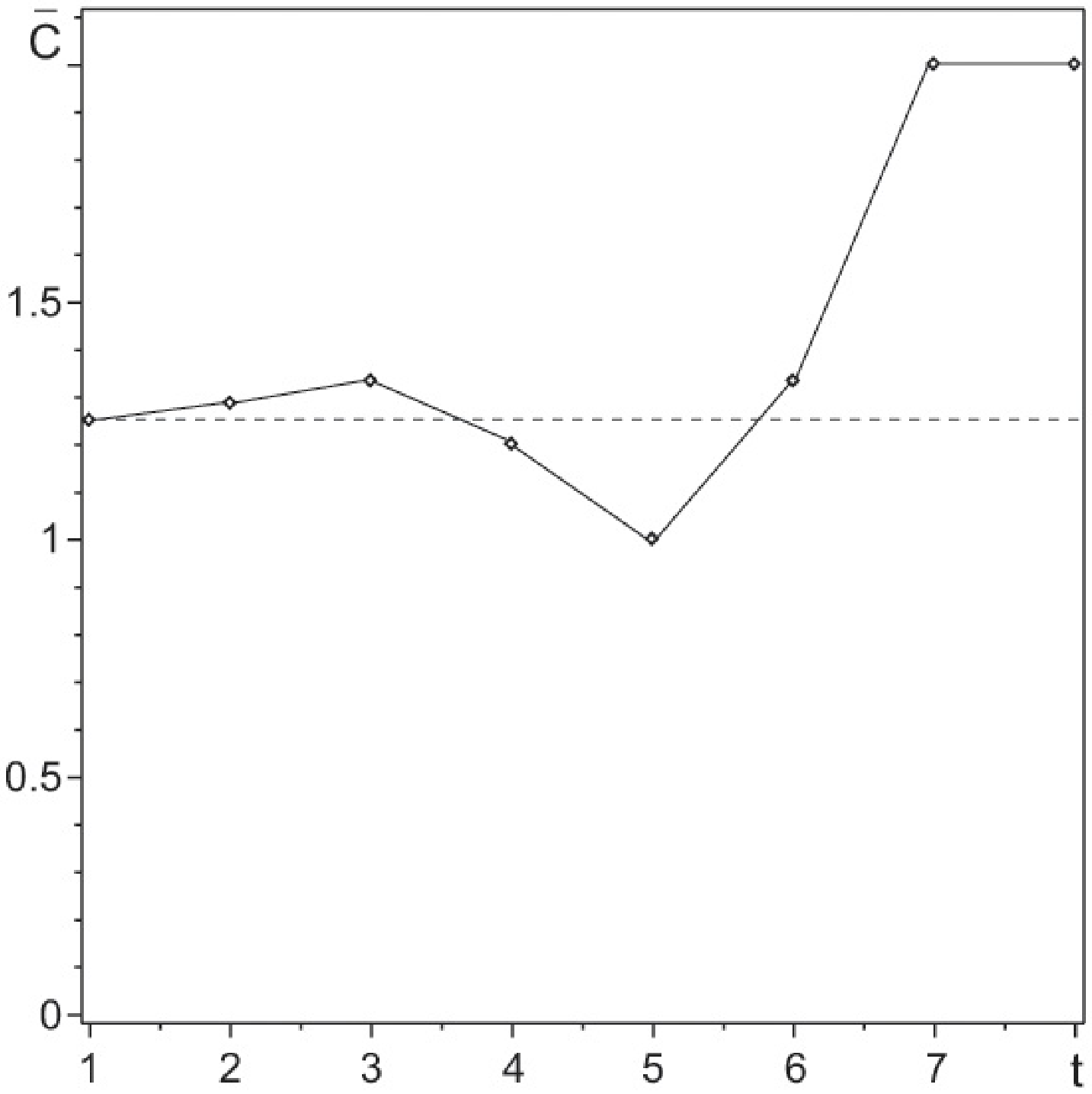}
  \includegraphics[width=5 cm]{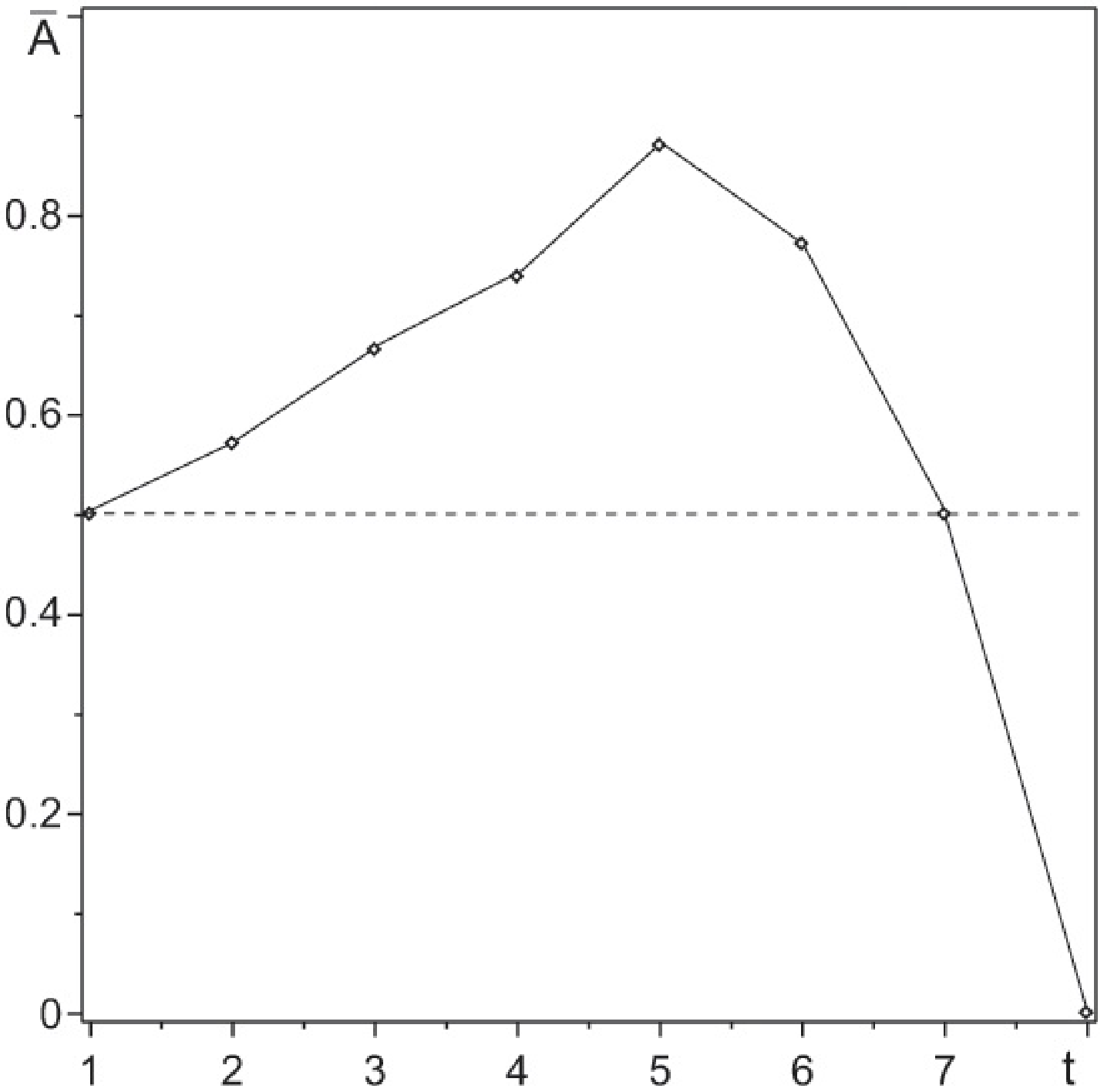}
  \includegraphics[width=5 cm]{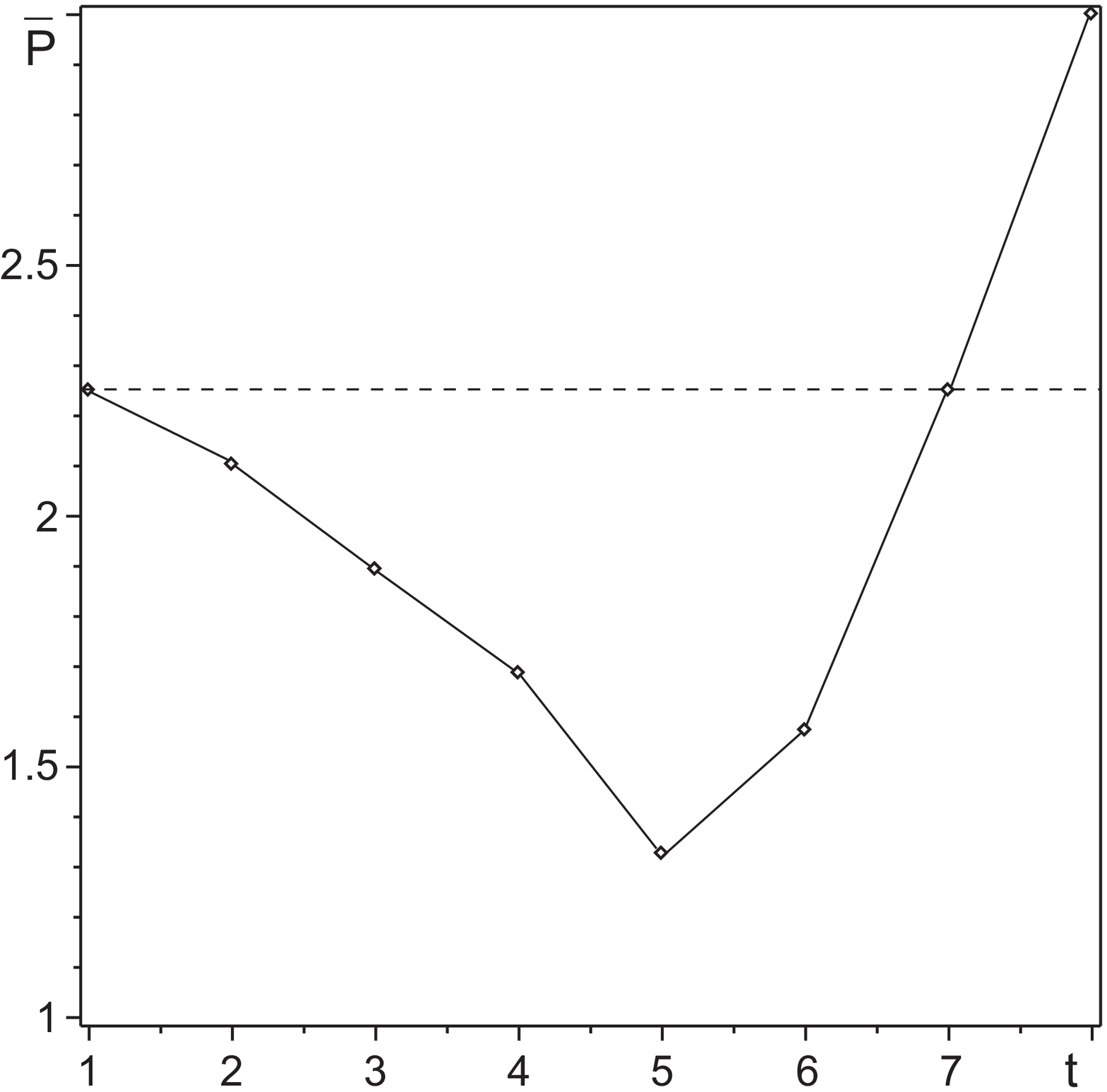}\\
  \caption{On the left is the change in the average value of the complexity of the whole "population" in the process of evolution. The dashed line is the initial average value of the complexity of "population". In the center change in the average aggressiveness of the "population" over time. The dashed line coincides with the aggressiveness of a population in which all strategies are present. In this case, the aggressiveness is 0.5. On the right is the change in the average number of evolutionary advantage points earned by the strategy in one move at each stage of evolution.}\label{fg5}
\end{figure}

We now turn to a discussion of changes in the complexity of social strategies. This is the main characteristic by which strategies in this world can be classified. The most detailed information on the behavior of complexity is carried by the number of strategies of corresponding complexity at each stage of evolution. In a world with zero memory, there are strategies of complexity 0, 1, and 2. Graphs of changes over time of the number of strategies of a certain complexity are shown in Fig.\ref{fg4}, where $n_0 (t)$ is the number of strategies of complexity 0 at the $t$-th stage of evolution, $n_1 (t )$ and $n_2 (t)$ are the number of strategies of complexity 1 and 2, respectively, at the $t$th stage of evolution. From these dependencies it is clear that strategies of complexity 1 disappear first at the 3rd stage of evolution. Strategies of zero complexity disappear only at the 7th stage of evolution. Actually, evolution ends here and the winning complex strategy with complexity 2 remains. In our case, this is strategy $[1] 01$. Accordingly, in the world without memory, the "eye for eye"(or the "tooth for tooth") strategy wins $[1] 01$. The complexity of the winning strategy is maximum in this class of strategies.

The dependencies shown in Fig.\ref{fg4} make it easy to obtain the average value of the complexity of the entire "population" at each stage of evolution. The average value of complexity is defined as
\[ \bar{C}(t)=\frac{0 \cdot n_0(t) + 1\cdot n_1(t) +2 \cdot n_2(t)}{ n_0 (t) +  n_1(t) + n_2(t)} \equiv \frac{ 1\cdot n_1(t) +2 \cdot n_2(t)}{ n_0 (t) +  n_1(t) + n_2(t)}\]
The dependence of the average complexity of the strategies of "population" on the time of evolution is shown in Fig.\ref{fg5}.

The average complexity of such a population demonstrates quite nontrivial behavior. At the beginning of evolution, the complexity of strategies increases slightly, but then decreases. After reaching a certain minimum value, the average complexity begins to increase to a maximum value. The population reaches a minimum of complexity at stage 5.

We now discuss which strategies win at different stages of the evolution of population or which strategies dominate population. We will monitor the complexity of the winning strategies at different stages of evolution.

\begin{figure}
  \centering
  \includegraphics[width=6 cm]{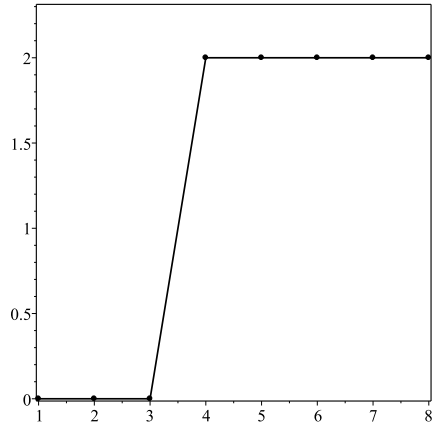}\\
  \caption{The complexity of the winning strategy at the appropriate stage of evolution.}\label{fg6}
\end{figure}

Figure \ref{fg6} shows the corresponding dependence. It can be seen that in the early stages of evolution (up to and including 3 stages) only primitive strategies with zero complexity won. After this stage, the most complex strategies win.

Observing the characteristic behavior of the complexity of the strategies of "population", one can divide the evolution time into two periods. The primitive period in which complexity decreases to a minimum and the period of a developed population of strategies in which the complexity of strategies increases. Then, the initial stage of evolution of the "population" of strategies can be described as a primitive world (up to stage 5 inclusive). The stage of primitive population in a world without memory lasts $62.5 \%$ of the time it takes to go to hospital. At these stages in the "population" primitive strategies of zero complexity dominate (see Fig.\ref{fg6}). The final stages correspond to a developed "population", where complex (even the most complex) strategies dominate.

However, it should be noted that primitive strategies are present in population even after the onset of the stage of a developed "population". The last primitive strategy disappears only at the 6th stage of evolution (this is strategy $(1)00$) (see Fig.\ref{fg4}).

Let us now consider another important characteristic of strategies. It can conditionally be called aggressive strategy. By aggressiveness we mean the share of strategy failures from cooperation. We will roughly determine it by the fraction $0$ in the name of the strategy. More precisely, we will determine it by the   average fraction of non-cooperation. Modeling the evolution of strategies gives a change in the average aggressiveness of population, shown in Fig.\ref{fg5}.

From this dependence it can be established that the primitive stage of the development of population is also characterized by an increase in average aggressiveness. At the end of the period of primitive population, its aggressiveness is maximum. Then, after the transition to a developed population, a monotonous decrease in the average aggressiveness of the population is observed, and when the stationary state is reached, the average aggressiveness is zero.

It is interesting to note that the most aggressive strategy also dies out at the 5th stage of evolution, the most "decent" at the first stage. Thus, it is possible to determine the primitive era of population by the growth of aggressiveness and the primitive stage ends after reaching the maximum aggressiveness of the strategies of population. In a world without memory, this gives an equivalent definition of the primitive era of population.

The third possibility of a reasonable definition of the primitive stage is associated with the period of the presence of the most aggressive strategy in population. Her disappearance marks a transition to a developed population.

Finally, we move on to discussing the set of evolutionary advantage points by strategies at different stages of evolution. This characteristic makes it possible to compare the set of points by strategies at different stages of evolution. As such a characteristic, you can use the number of points scored on one course of the strategy on average at a certain stage of evolution. The time dependence of this value is shown in Fig.\ref{fg5}. It is easy to see that with increasing aggressiveness, the average number of points that a strategy gains decreases. The higher the aggressiveness, the lower the number of points scored. At the stage of a developed population, the obtained number of points begins to grow monotonously, reaching a maximum at the stationary stage. Comparing the aggressiveness and the average number of points that the strategy scores in Figure 5, it is easy to notice a correlation between the behavior of these characteristics during evolution.
\begin{figure}[t]
  \centering
  \includegraphics[width=6 cm]{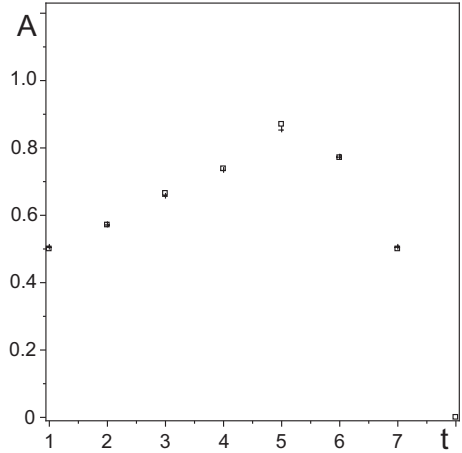}\\
  \caption{Comparison of the average aggressiveness obtained by numerical modeling (squares), with the ratio (\ref{q1}) (crosses). The difference is noticeable only at the maximum, the rest is superimposed dots.}\label{fg7}
\end{figure}

In a world with zero memory, there is a correlation in the behavior of average aggressiveness and average earnings per move. It can be assumed that the relationship between these characteristics is determined by the relation
\begin{equation}\label{q1}
  \bar{A}(t)=\sqrt{\lambda \cdot (\bar{P}_{max}-\bar{P}(t))}-a
\end{equation}
Figure\ref{fg7} compares the average aggressiveness obtained by numerical simulation with the empirical pattern given above. The scale factor was chosen for reasons of the equality of these characteristics at the first stage of evolution, $\lambda = 5.3/8$ and $a=0.2$. Despite a slight deviation in the maximum region, the graphs show good agreement in the behavior of these characteristics over time.

Such a relationship (\ref{q1}) establishes that a decrease in the number of points per move leads to an increase in the aggressiveness of population. Naturally, we can rewrite relation (\ref{q1}), resolving it with respect to $\bar{P}(t)$. Then it can be argued that an increase in aggressiveness leads to a decrease in the average number of points per strategy course according to $\bar{P}(t) =\bar{P}_{max} - \frac{(\bar{A}(t)+a)^2}{\lambda}$.
Thus, the number  of points per strategy move quadratically depends on aggressiveness. We now turn to strategies with minimal memory and analyze the change in the behavior of strategies in the process of evolution.

\section{The world with a depth of memory 1}

Let's move on to the world of strategies with a depth of memory 1. In this world, the number of all strategies increases and becomes equal to 128 (see section 2). It is clear that tracking each strategy, although still possible, is becoming less meaningful. Such detailed information is more confusing than helping to understand the patterns of behavior of strategies. Therefore, with increasing memory depth and, accordingly, the number of strategies, a collective way of describing strategies becomes extremely important. In a world with a memory depth of 1, strategies differ in complexity (0, 1, 2, 3, 4) and also in memory depth ($0,1$). These characteristics make it possible to classify all strategies into groups according to these properties. Thus, in this world, in addition to the characteristics of the number of strategies of a certain complexity ($n_0$, $n_1$, $n_2$, $n_3$ and $n_4$), other characteristics can be introduced -- the number of strategies of a certain memory depth ($a_0$, $a_1$). These are very important collective variables that allow you to describe the properties of a large number of different strategies. In the process of evolution, these numbers change and give an abbreviated description of the behavior of strategies. The evolution time in this world is 100. There are 4 strategies left.
\begin{figure}
  \centering
  \includegraphics[height=5 cm]{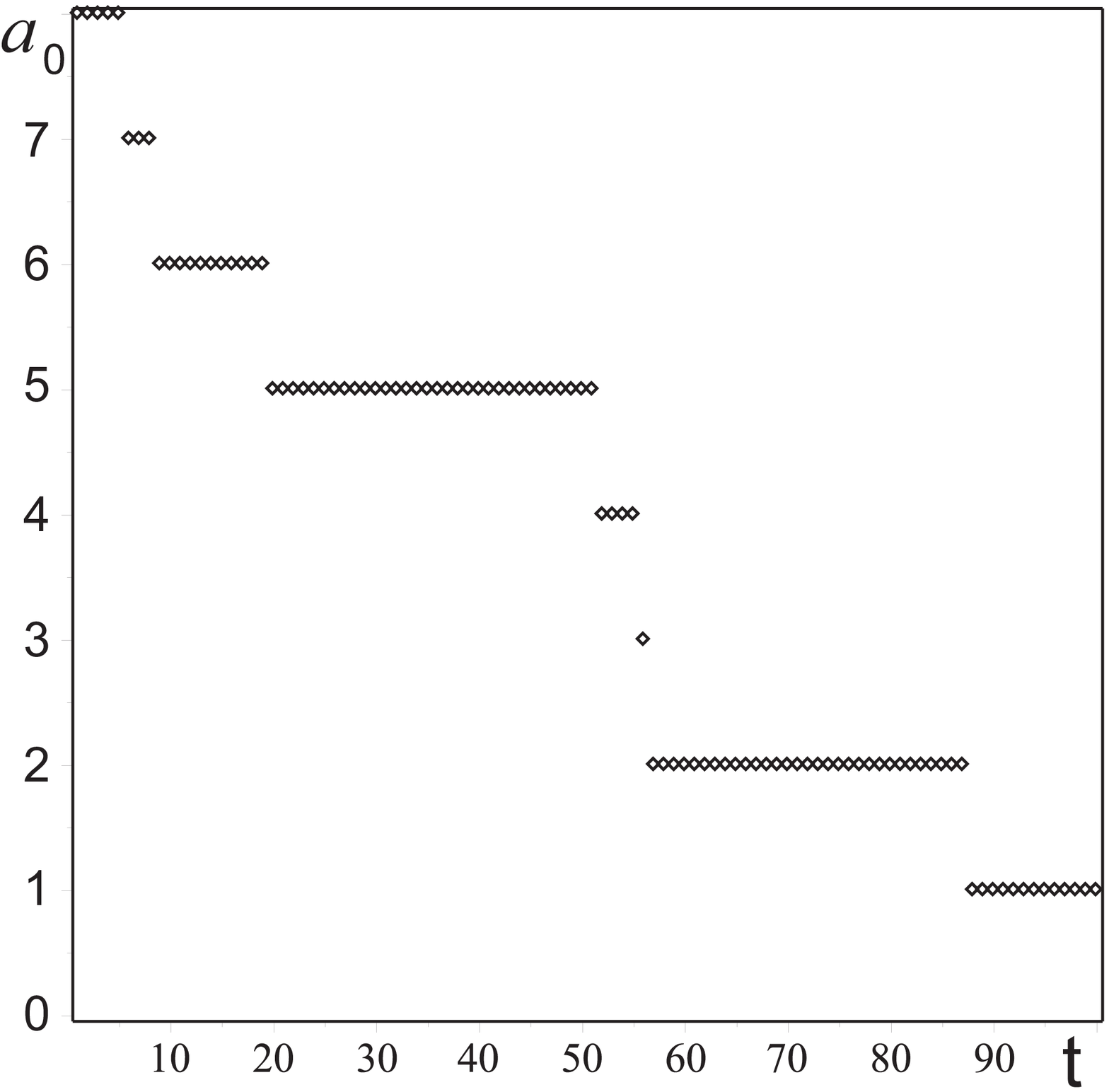}\
  \includegraphics[height=5 cm]{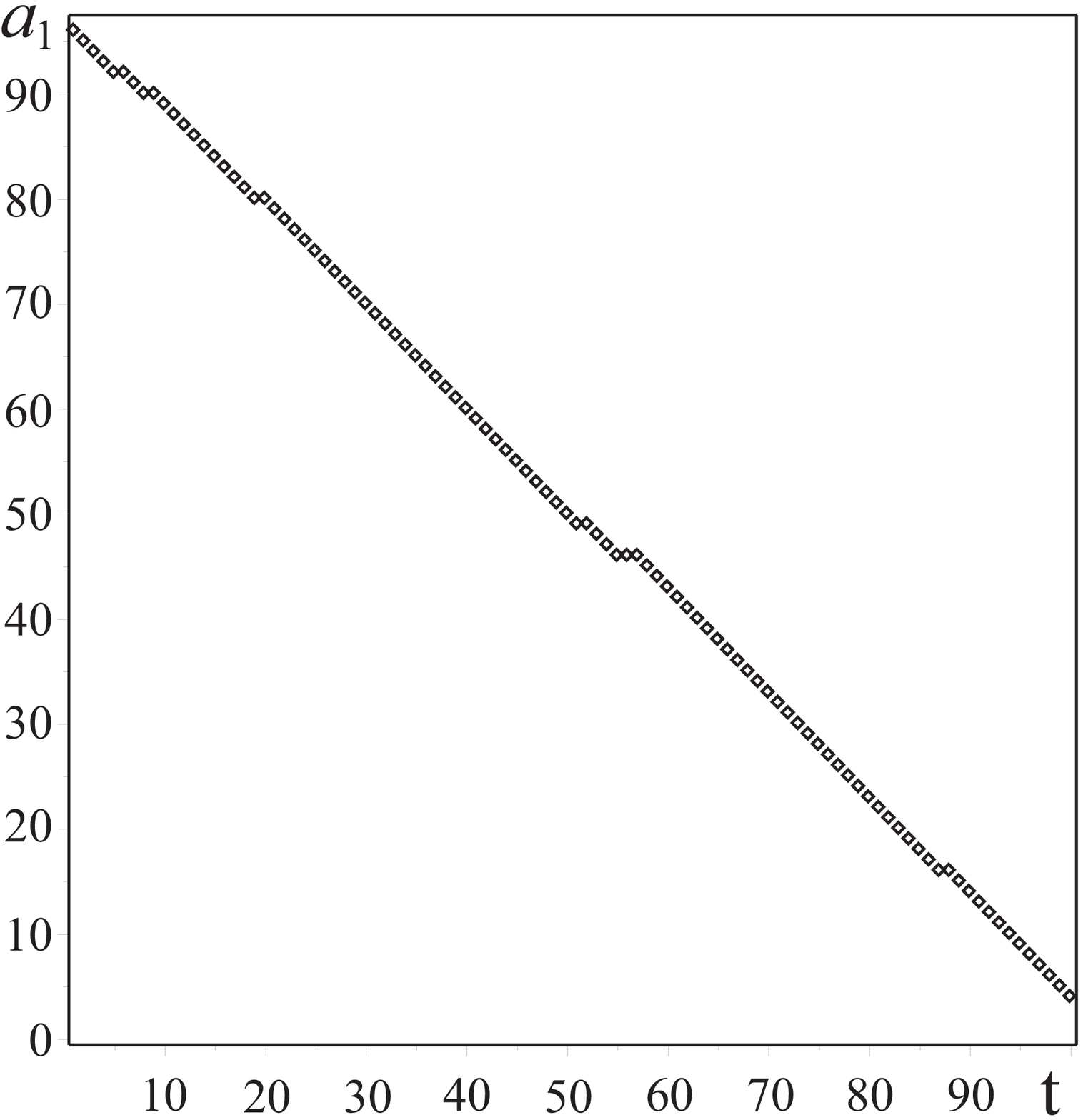}\\
  \caption{On the left is the dependence of the number of strategies with zero memory depth, on the right with a depth of 1 on time.}\label{fg8}
\end{figure}

We start by discussing memory changes during evolution. The most complete information about this process can be extracted by observing the behavior of $a_0 (t)$ and $a_1(t)$. Of course, some patterns are associated with an exponential difference in the number of strategies of different memory depths. So, the initial number of strategies of the 0th memory depth is 8, and the depth of 1 is 96. Therefore, the discreteness of the change in the number of zero memory strategies is so noticeable in Fig.\ref{fg8}. The behavior of $a_1(t)$, although it looks like a linear function, has important differences from it. In addition, such a significant difference in numbers makes it uninformative to compare their behavior on the same graph. Other characteristics need to be used to compare their behavior. Fig.\ref{fg8} shows that strategies with zero memory depth are present in this world throughout the entire evolutionary period.

\begin{figure}
  \centering
  \includegraphics[width=5 cm]{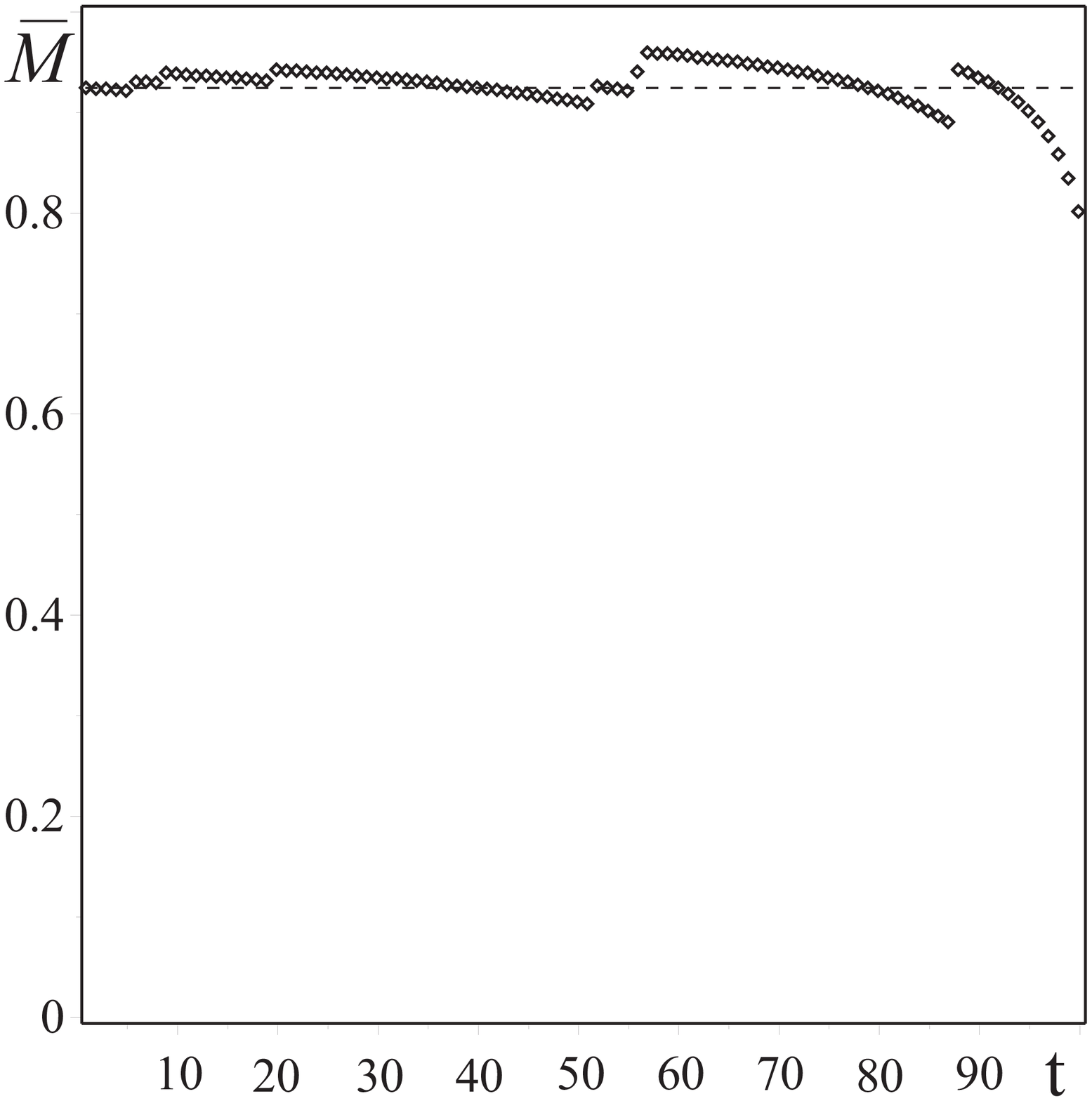}
  \includegraphics[width=5 cm]{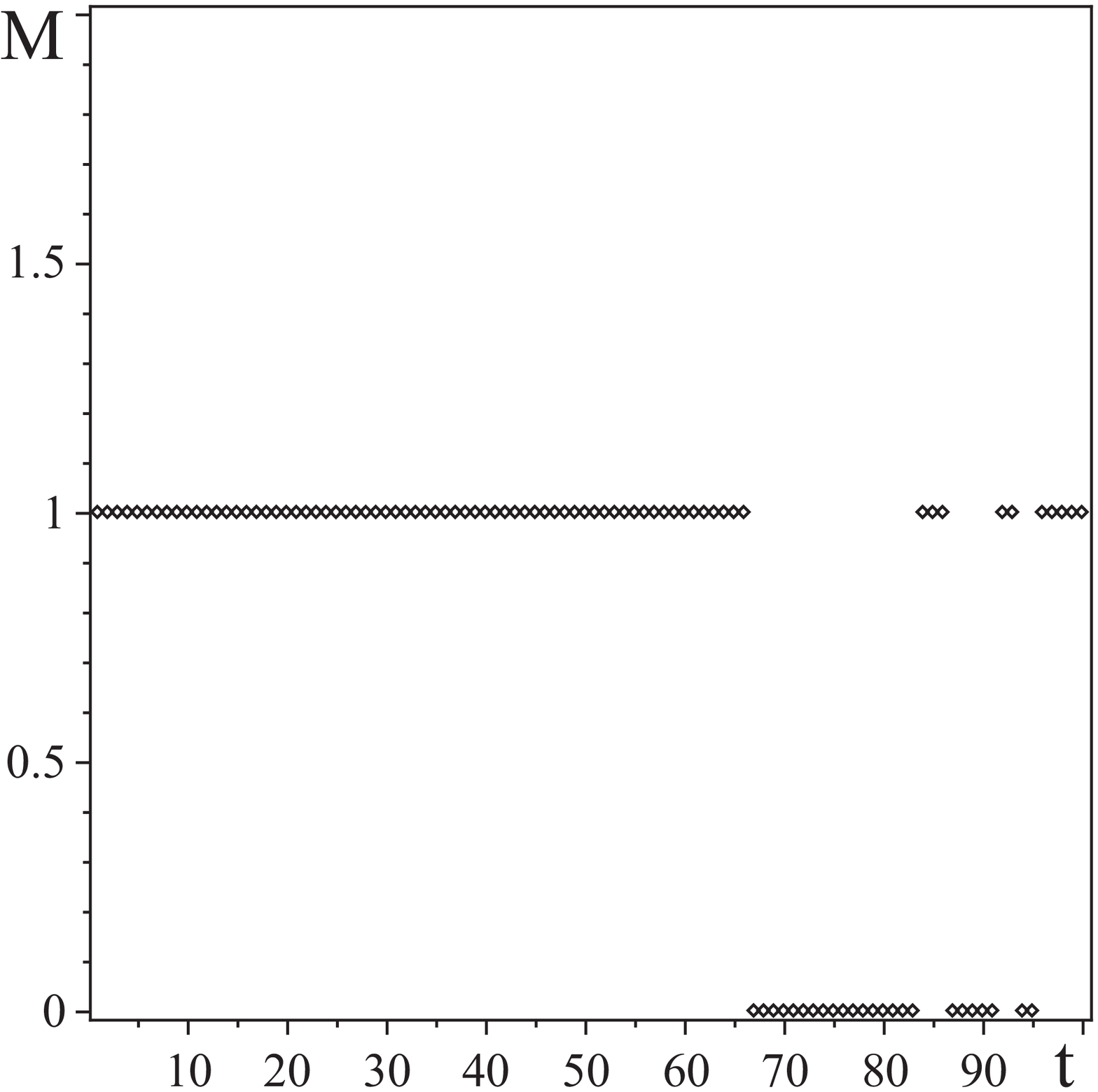}\\
  \caption{On the left is the change in the average memory of population. On the right is the memory depth of the winning strategies at the respective stages.}\label{fg9}
\end{figure}

The functions $a_0 (t)$ and $a_1(t)$ allow us to calculate the average depth of population's memory, which is defined as
\[\bar{M}=\frac{0\cdot a_0 + 1\cdot a_1}{a_0 +a_1}\equiv \frac{ a_1}{a_0 +a_1}\]
The result of averaging is shown in Fig.\ref{fg9}. Based on Fig.\ref{fg9}, it can be noted that in the process of evolution, the average depth of population's memory changes insignificantly. The reason for this is due to the small number of strategies with zero memory and the presence of strategies with a greater depth of memory, even when entering a stationary state. Figure\ref{fg9} shows the memory depths of the winning strategies at the corresponding stages of evolution. It is easy to see that in the initial period, the dominant strategies have a maximum memory depth. In this case, periods may arise when the dominant strategy has a shallow depth of memory. However, this does not have a significant impact on the average memory of population. This is a consequence of the relatively small number of strategies with small memory, even at these stages.

\begin{figure}
\centering
  \includegraphics[height=5 cm]{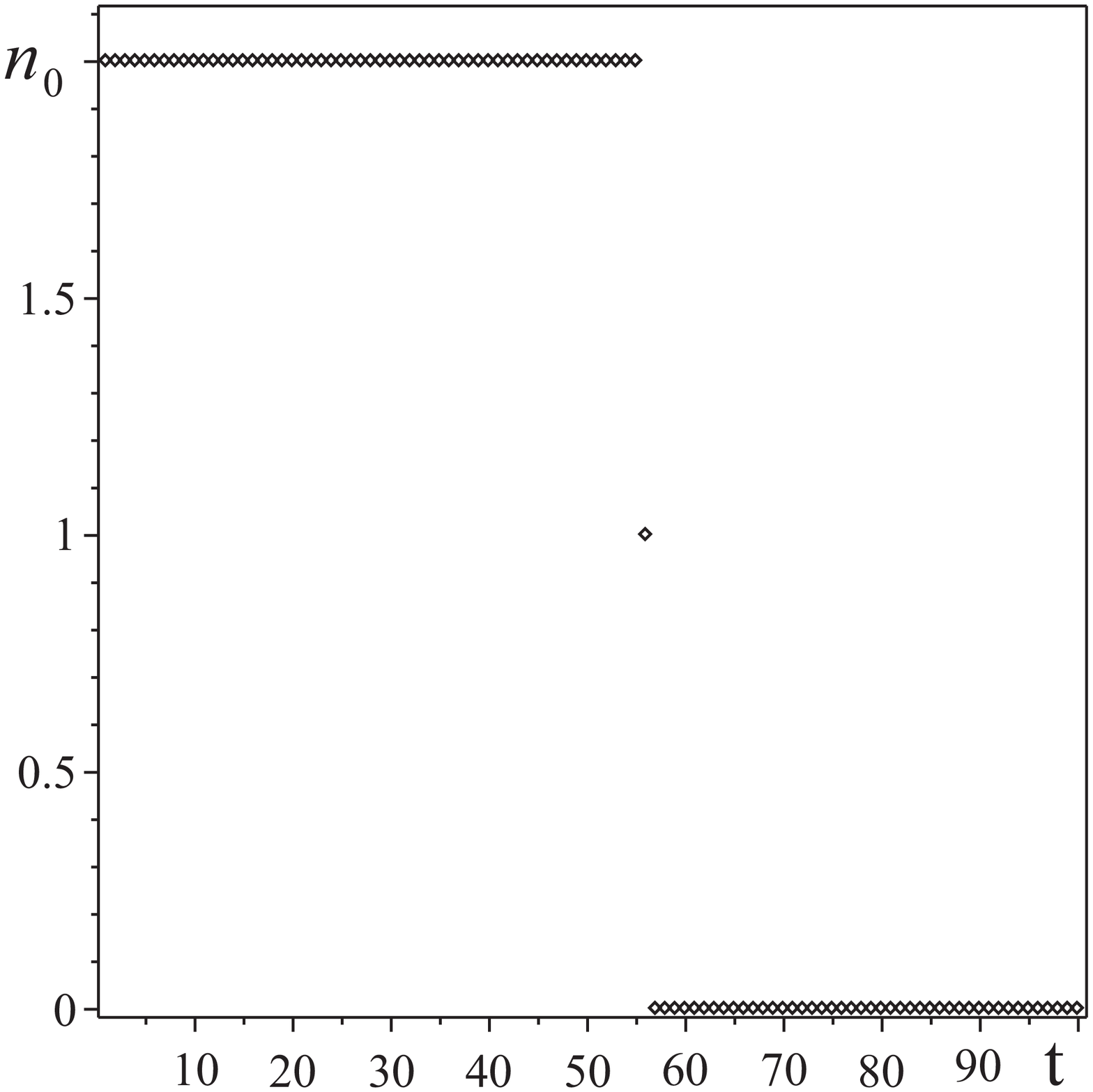}
  \includegraphics[height=5 cm]{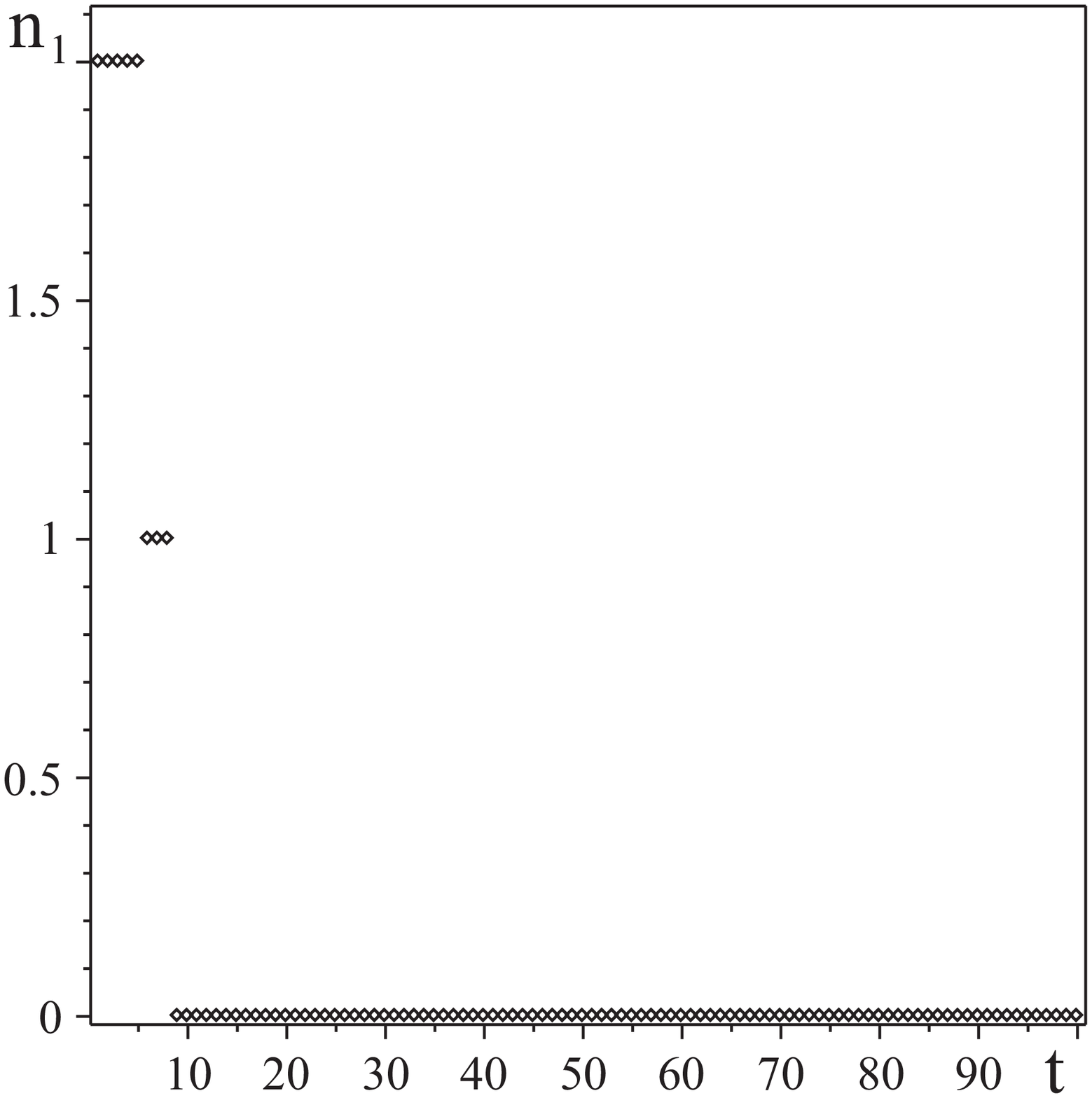}
  \includegraphics[height=5 cm]{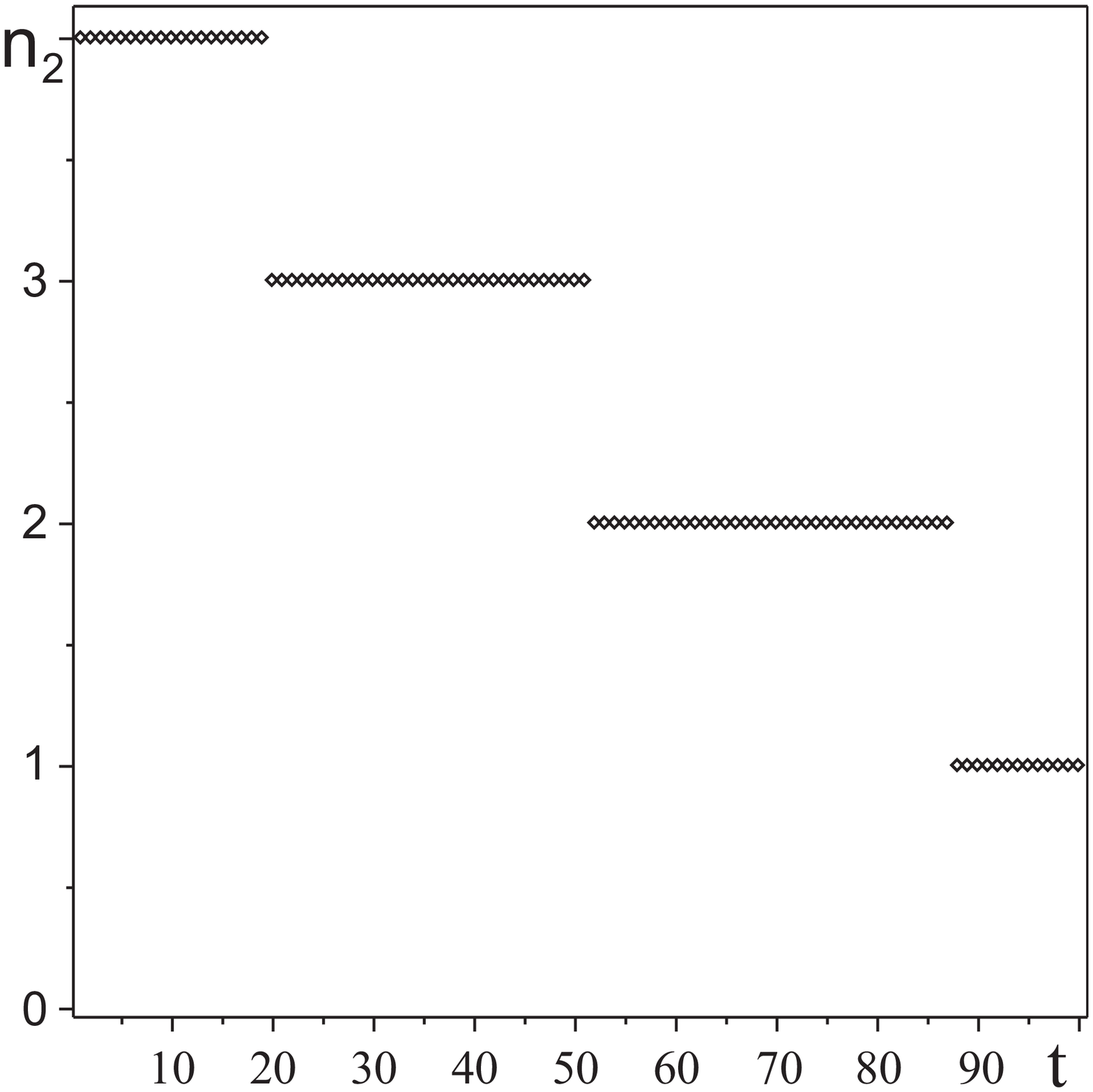}
  \includegraphics[height=5 cm]{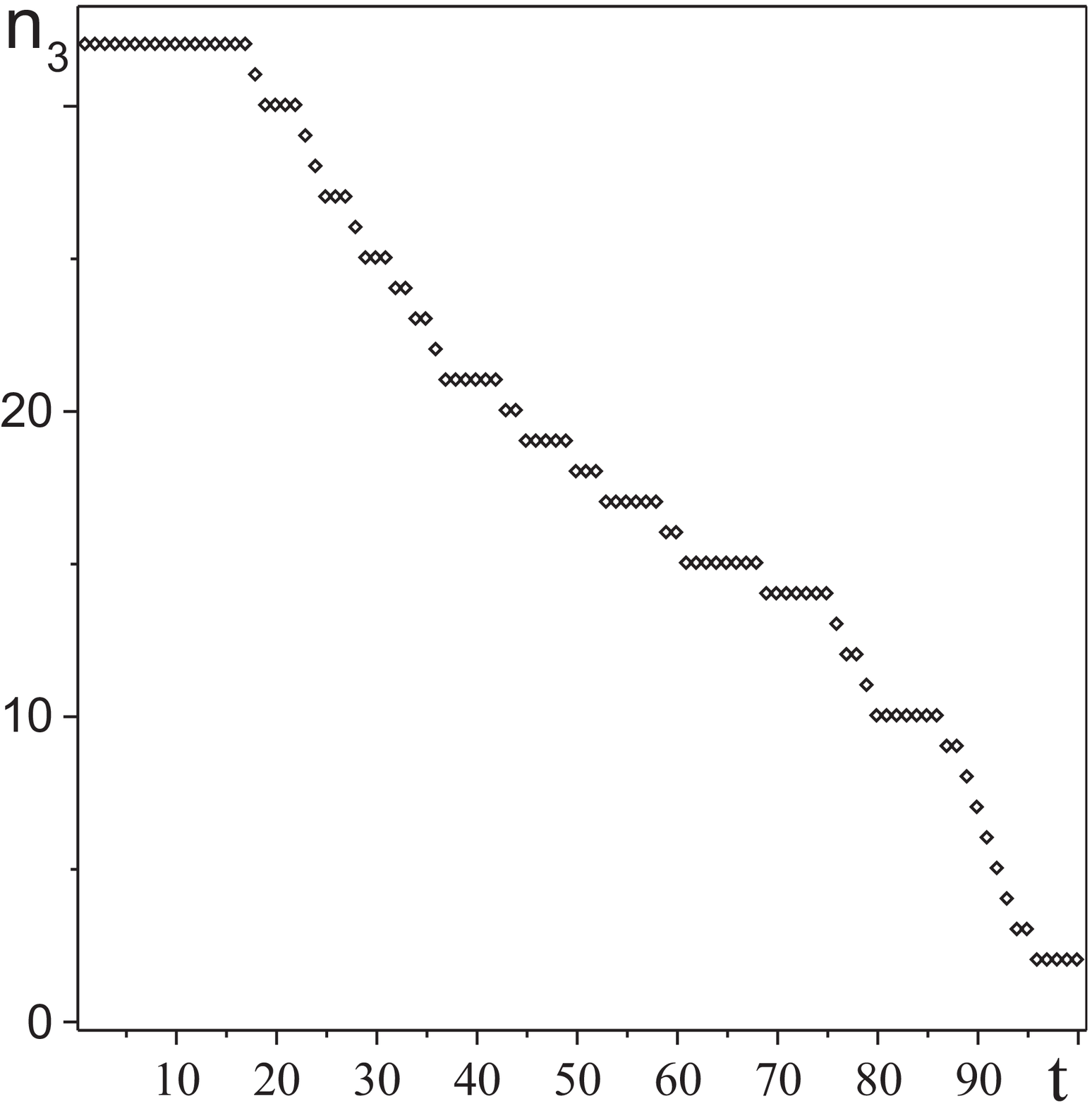}
  \includegraphics[height=5 cm]{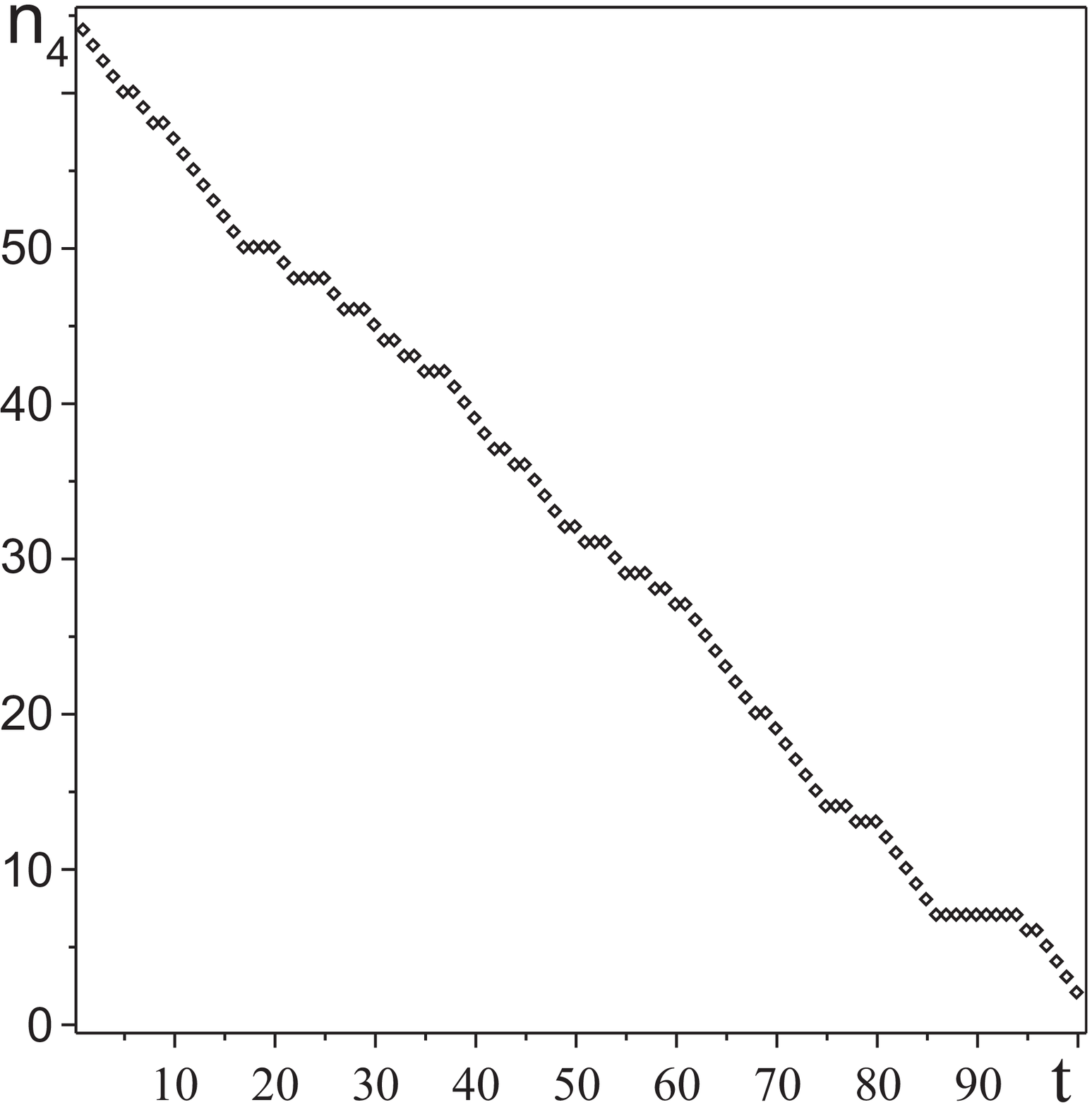}\\
  \caption{The number of strategies of corresponding complexity present in population at different times of evolution.}\label{fg10}
\end{figure}

We turn to the analysis of the behavior of the complexity of population. The most detailed information on the complexity of population is provided by the functions $n_0(t)$, $n_1(t)$, $n_2(t)$, $n_3(t)$  and $n_4(t)$. These characteristics are shown in Fig.\ref{fg10}. Let us pay attention to the initial stage, in which the most primitive strategies $n_0 (t)$ are present. This period takes 55 stages of evolution and ends after the most aggressive strategy $0000$ disappears. Unlike strategies with 0-th memory in this world, primitive strategies do not dominate this period. However, they are present in population in full force. Therefore, the primitive period of development of a population with memory is not characterized by the dominance of the most primitive strategies, but is determined by their presence. The primitive period takes $55 \%$ of the time to reach the stationary state. Note that the relative duration of the primitive period decreased with an increase, or more precisely, with the advent of memory.

The very first strategies to disappear from population are those with complexity 1 (see Fig.\ref{fg10} - dependence $n_1 (t))$, which include the most respectable strategy $1111$, which disappears already at stage 5. One can only be surprised that it did not disappear at the first stage of evolution.
It is interesting to note that the average rate of disappearance of strategies is the highest for the most complex strategies. The higher the complexity, the greater the average rate of disappearance of the corresponding strategies. Despite this, complex strategies survive.

\begin{figure}
  \centering
  \includegraphics[width=5 cm]{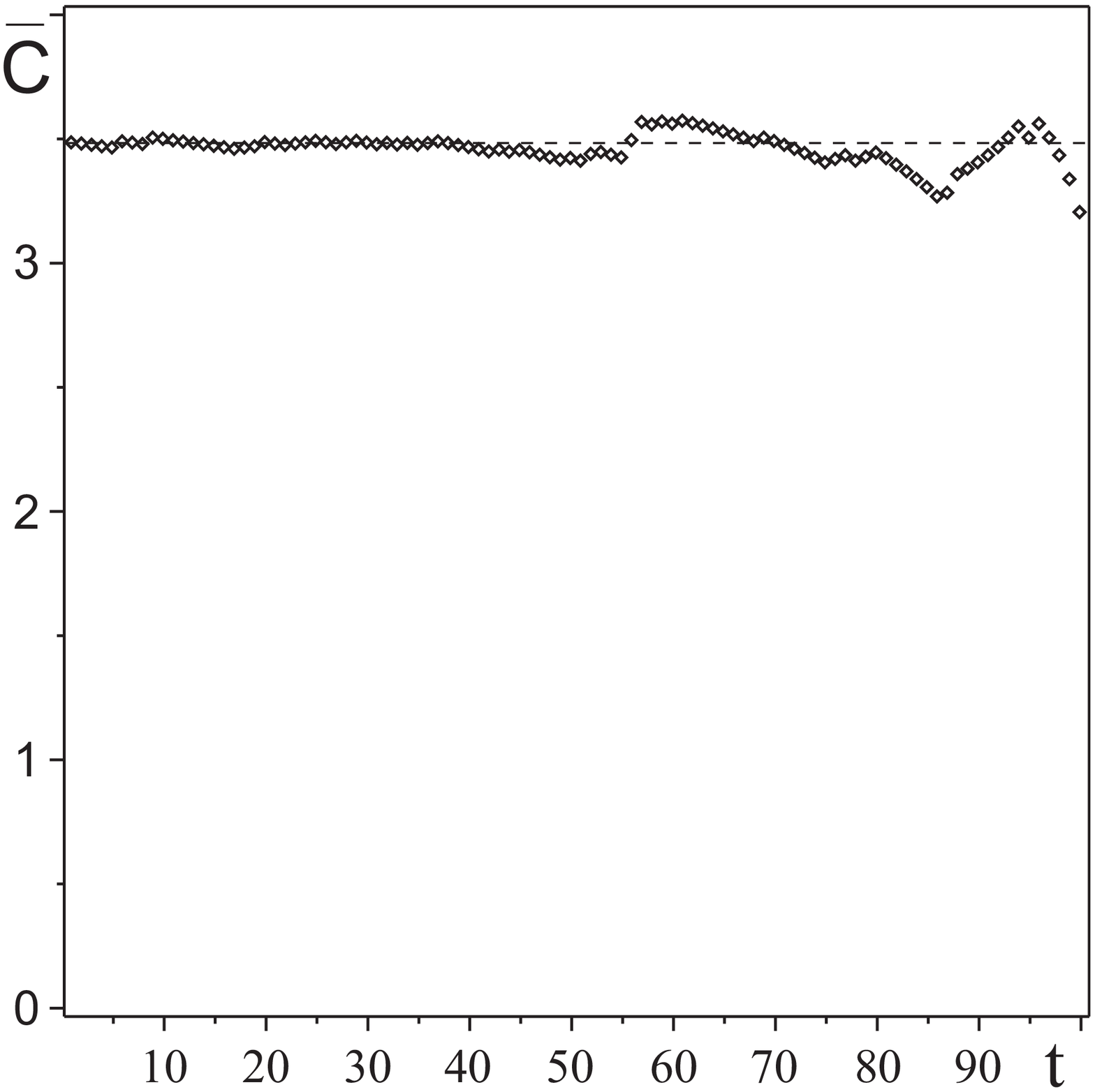}
  \includegraphics[width=5 cm]{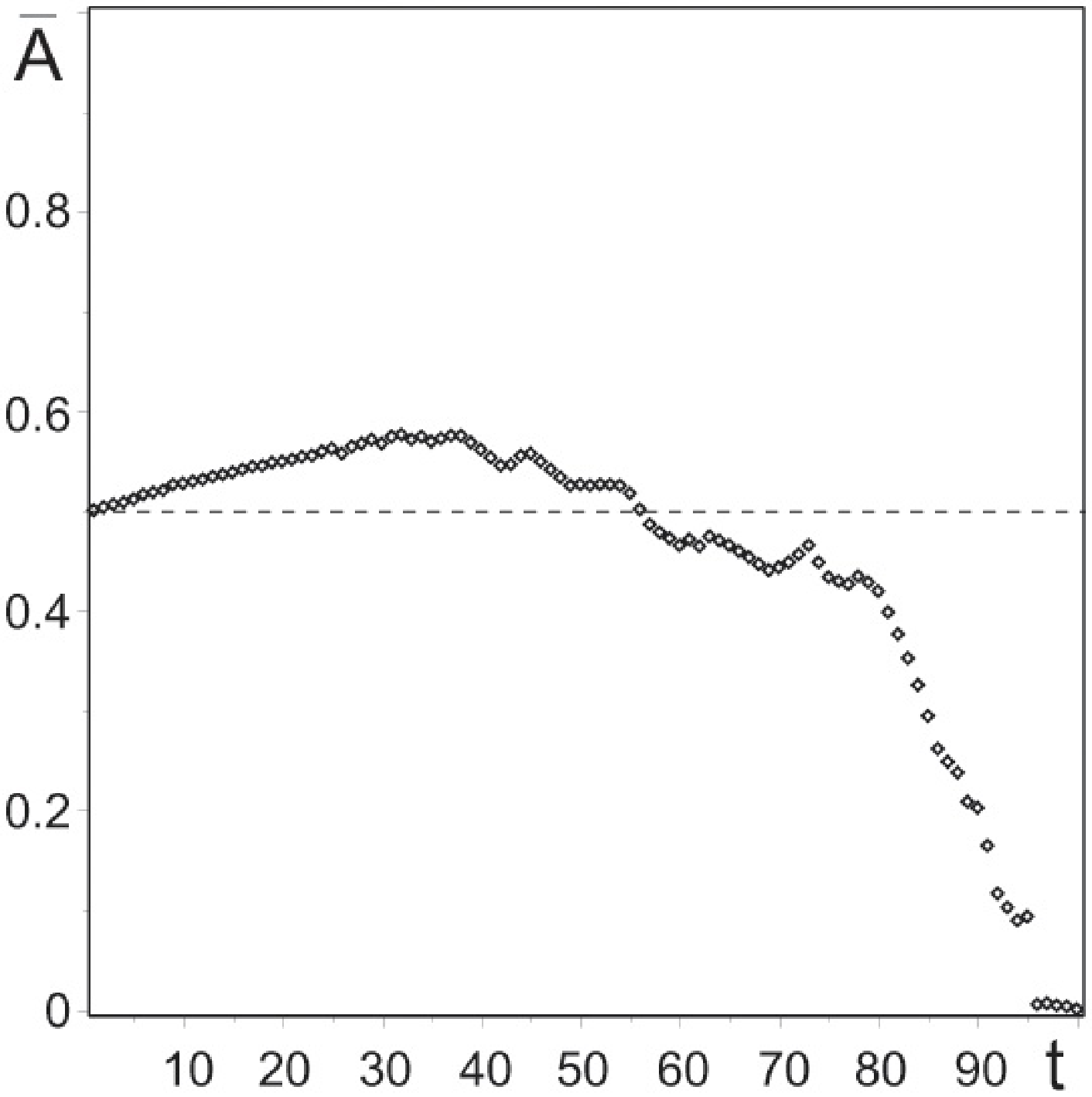}
  \includegraphics[width=5 cm]{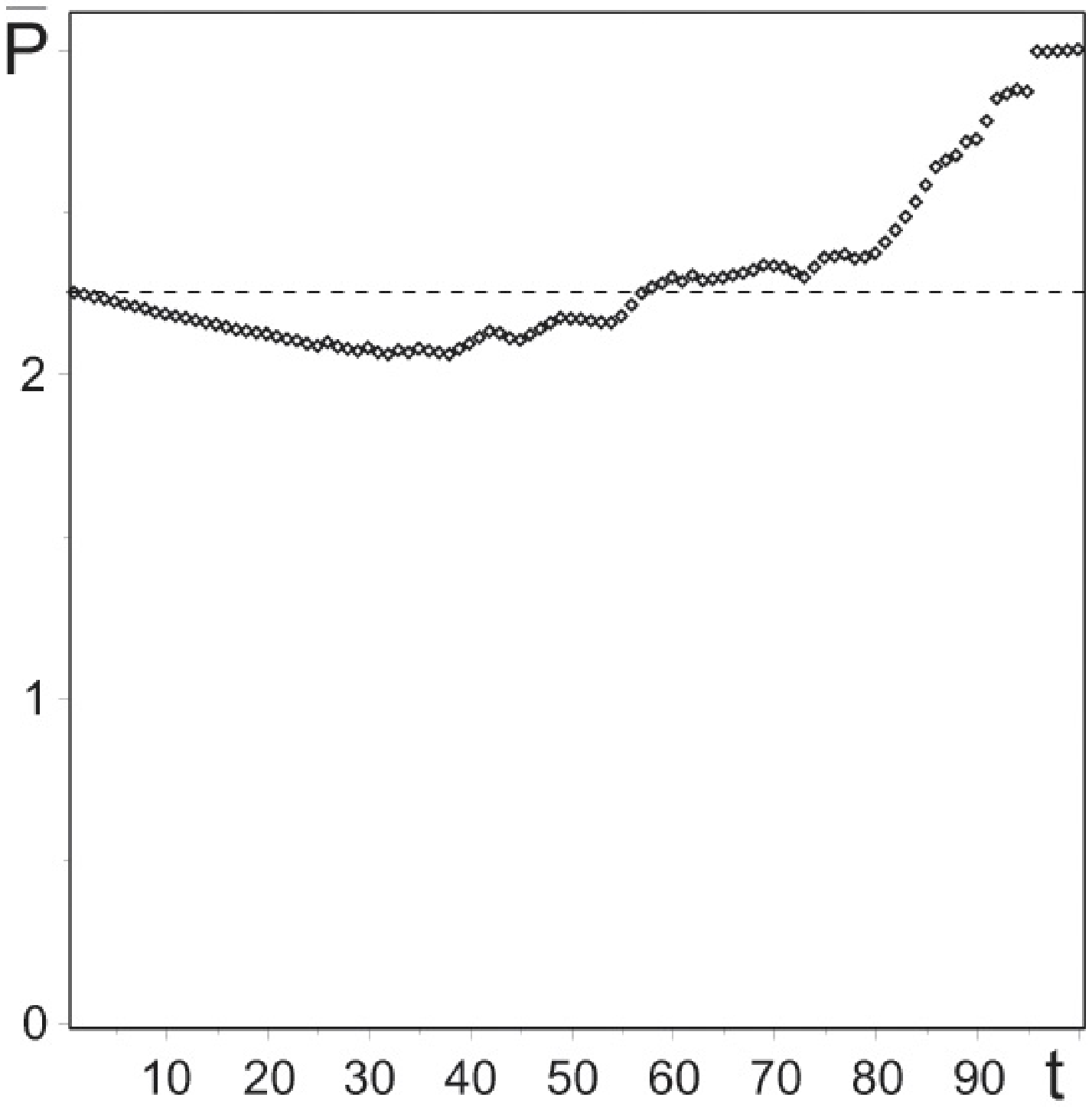}\\
  \caption{On the left is the change in the average complexity of social strategies over time. In the center is the change over time of the average aggressiveness of population with a depth of memory of 1. On the right is the average earnings per move at each stage of evolution. The dashed line shows the corresponding characteristics of a population in which all strategies are present with depth memory 1 and below.}\label{fg11}
\end{figure}

This is clearly seen in the change in the average complexity of population's strategies in the process of evolution (see Fig.\ref{fg11}). It can be seen that the average complexity varies slightly

\begin{figure}
  \centering
  \includegraphics[width=5 cm]{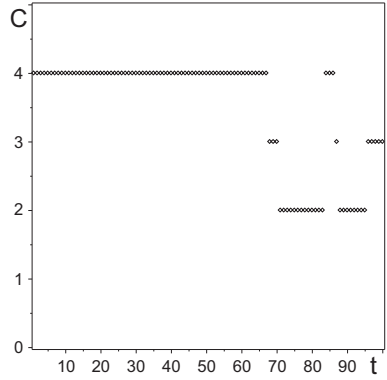}\\
  \caption{The complexity of winning strategies at different stages of evolution. There is no stage of dominance of primitive strategies. At all stages of evolution, the most complex strategies or those close to maximum complexity dominate.}\label{fg12}
\end{figure}

In addition, you can see that in a world with a memory depth of 1, complex strategies dominated at all stages of evolution. This is clearly seen from Fig.\ref{fg12}, which indicates the complexity of the winning strategies at each stage of the evolution of population. It is clearly seen that the most complex strategies capture the primacy from the beginning of evolution and retain it, or rather, sharing it with strategies that are close in complexity to the maximum.
Let us now examine how the aggressiveness of strategies changes in the process of evolution. Figure\ref{fg11} shows the change in the average aggressiveness of population. It is easy to see that, as in the previous world, the average aggressiveness at the initial times grows and exceeds the average aggressiveness of a population in which all strategies are present. Then the aggressiveness decreases and reaches the average aggressiveness of a population in which all strategies are present at the 55-56 stage of evolution. Further, the level of aggressiveness continues to decline, reaching a minimum when entering a stationary state. Qualitatively, this behavior resembles aggressive behavior in the absence of memory. The difference lies in the shift of the maximum in the presence of memory in relatively earlier times of evolution. So, the position of the maximum with a memory depth of 1 is reached at times making up $37 \%$ of the evolution time, and in the absence of memory - at times $62.5 \%$ of the evolutionary time. Thus, if you define a primitive period to achieve the average aggressiveness of the maximum, then the stage of primitive population ends at 37, 38 stages. The stage of primitive population with a depth of memory of 1 is significantly shortened and makes up $37 \% - 38 \%$ of the time of evolution.

Now let's look at how the set of evolutionary advantage points is changing. Figure\ref{fg11} shows the number of points on average per turn at each stage of evolution. As in the world with zero memory, there is a correlation in the behavior of average aggressiveness and average earnings per move. Assuming that the relationship between these characteristics is determined by the relation
\[\bar{A}(t)=\sqrt{\lambda \cdot (\bar{P}_{max}-\bar{P}(t))}-a \, ,\]
then we can compare the average aggressiveness obtained by direct modeling (see Fig.\ref{fg11}) with the aggressiveness obtained by average payouts per move.
\begin{figure}
  \centering
  \includegraphics[width=5 cm]{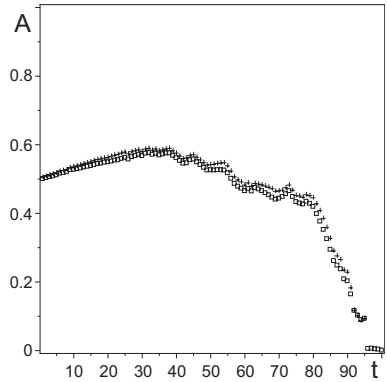}\\
  \caption{Comparison of average aggressiveness - squares with regularity (\ref{q1}), constructed according to the dependence of payments per move - crosses}\label{fg13}
\end{figure}
The scale factor $\lambda$ was chosen $\lambda= 5.3/8$ and $a=0.2$ so that the value at the first stage of these dependences coincided.

Such a comparison is shown in Fig.\ref{fg13}. It is easy to notice the proximity of the obtained dependencies. Of course, the rule is obtained empirically and the mechanism of such a connection is not entirely clear. However, good agreement between the same functions is observed with zero memory. In other words, these characteristics are not independent, and one of them depends on the other.

\section{The world with a depth of memory 2}

\begin{figure}
  \includegraphics[height=5 cm]{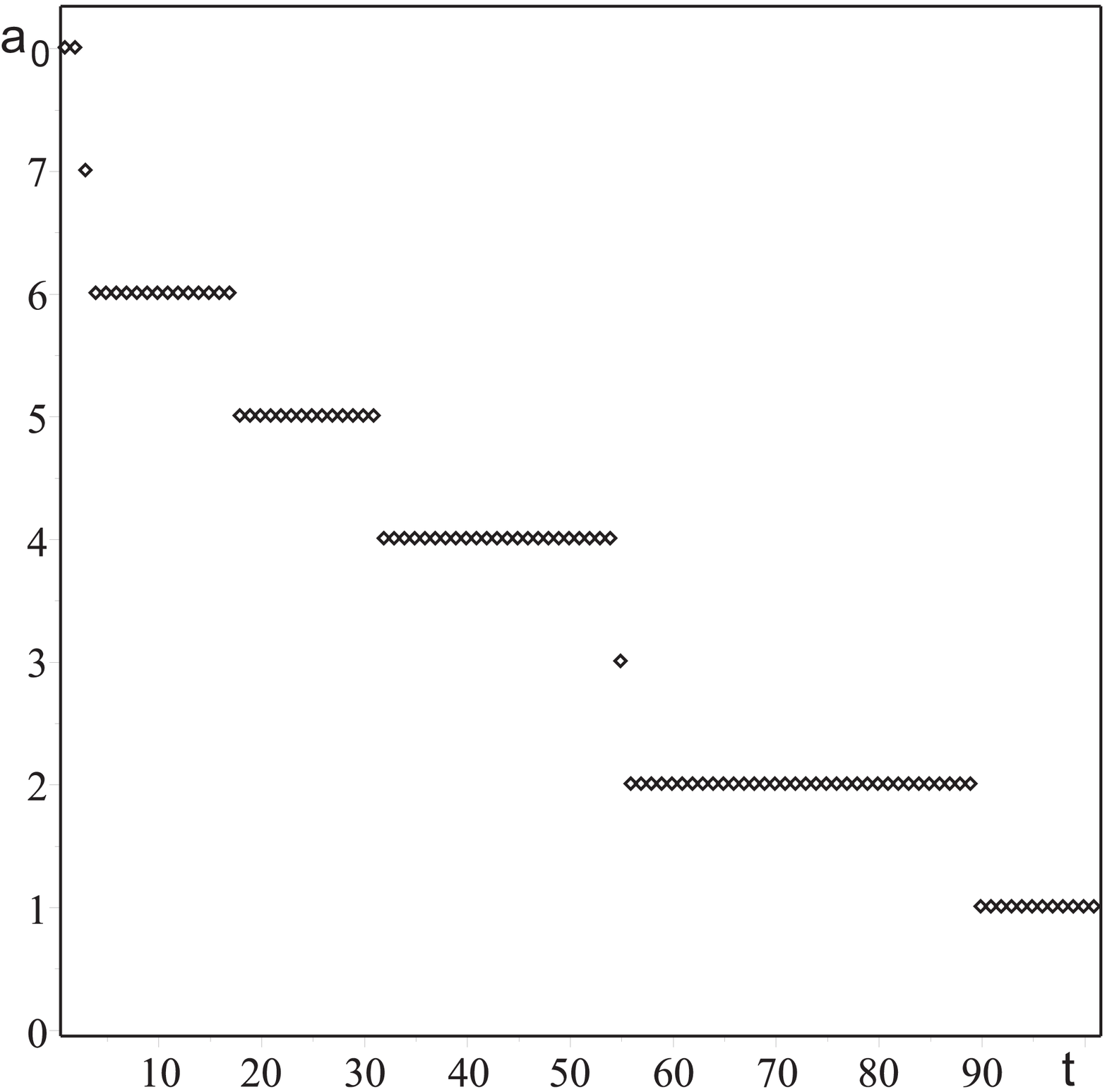}
  \includegraphics[height=5 cm]{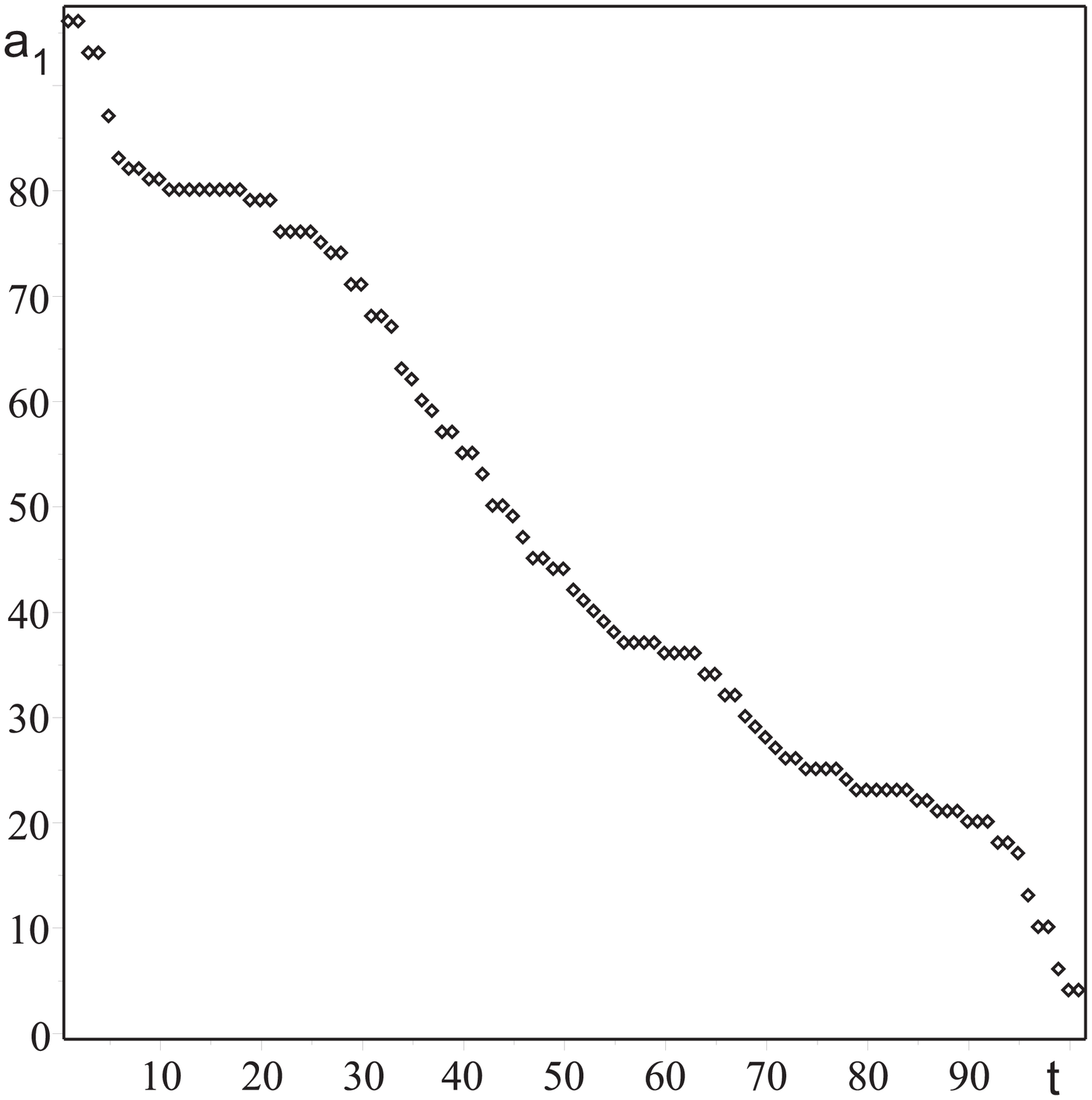}
  \includegraphics[height=5 cm]{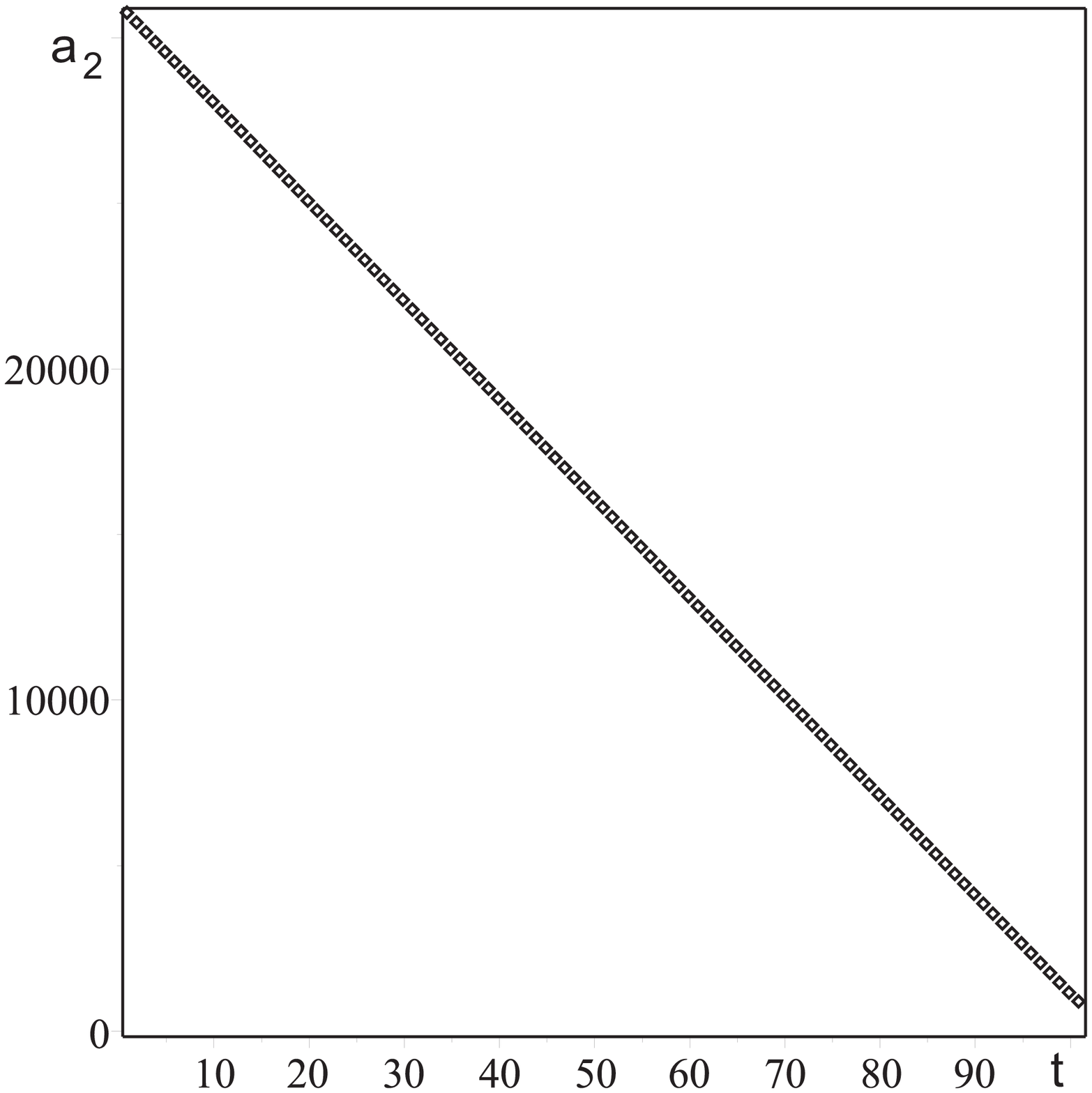}\\
  \caption{Change in the number of strategies with a certain depth of memory in the world with a depth of memory 2. A discrete unit of time is chosen equal to a duration of 300 generations.}\label{fg14}
\end{figure}

We now turn to the analysis of patterns in the world with a depth of memory 2. The payout matrix and the number of moves of the two strategies remain the same. Naturally, the number of all possible strategies in this world is increasing and is equal to 30824. Here again, a separate strategy is understood as a strategy with certain initial moves. As before, at each stage, the strategy (or strategies) that has gained the minimum number of evolutionary advantages will be deleted. Further everywhere, as a unit of time scale, we will use a duration of 300 generations. The time to reach the hospital is 29968 generations or the selected scale of 100. In terms of collective variables for analyzing the behavior of strategies with different memory depths, we will divide all strategies into 3 groups according to memory depth and we will monitor the changes in the numbers of these groups. So $a_0 (t)$ is the number of strategies in a population with a memory depth of 0 at the $t$-th stage, $a_1(t)$ is the number of strategies with a memory depth of 1 at the $t$-th stage, and $a_2 (t)$ is the number of strategies with a depth memory 2 at the $t$-th stage. When modeling the evolution of such a population, the dependence of the change in the number of these groups over time was obtained, which are shown in Fig.\ref{fg14}.

As expected, the discreteness of change is most pronounced for groups with a small memory depth of $0,1$ and is almost invisible for $a_2 (t)$. The reason for this is the relatively small number of strategies with a small depth of memory.

A characteristic feature that is clearly visible from Fig.\ref{fg14} is the presence of strategies with a shallow depth of memory for almost the entire time of evolution. In a certain sense, this is surprising when one considers the relatively small number of strategies with a small depth of memory. So, with zero memory depth, strategy $[1]01010101$ reaches the hospital, as well as strategies $[1][01]01110111$, $[1][11]01110111$ with memory depth 1 are included in the stationary set of strategies formed during evolution.

Using the functions $a_i(t)$, we consider how the average memory of population changes in the process of evolution.
\[\overline{M}=\frac{0 \cdot a_0 (t)+1 \cdot a_1 (t)+ 2 \cdot a_2 (t)}{ a_0 (t)+a_1 (t)+a_2 (t)}\]
The average memory depth of the strategy community is 2 and remains virtually unchanged over time. Apparently, this property will be preserved for strategy communities with a greater
depth of memory.
\begin{figure}
 \includegraphics[height=4.5 cm]{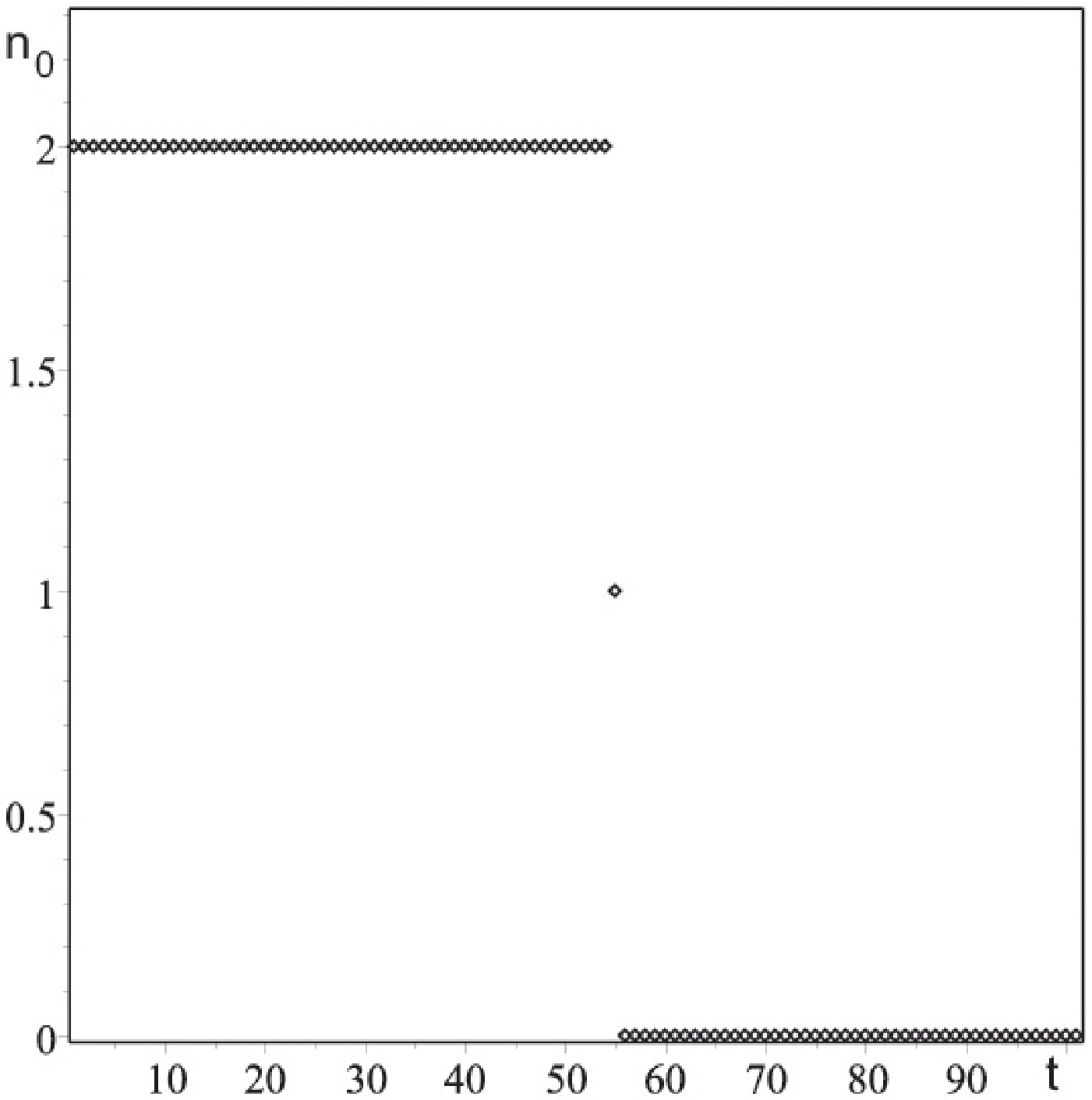}
  \includegraphics[height=4.5 cm]{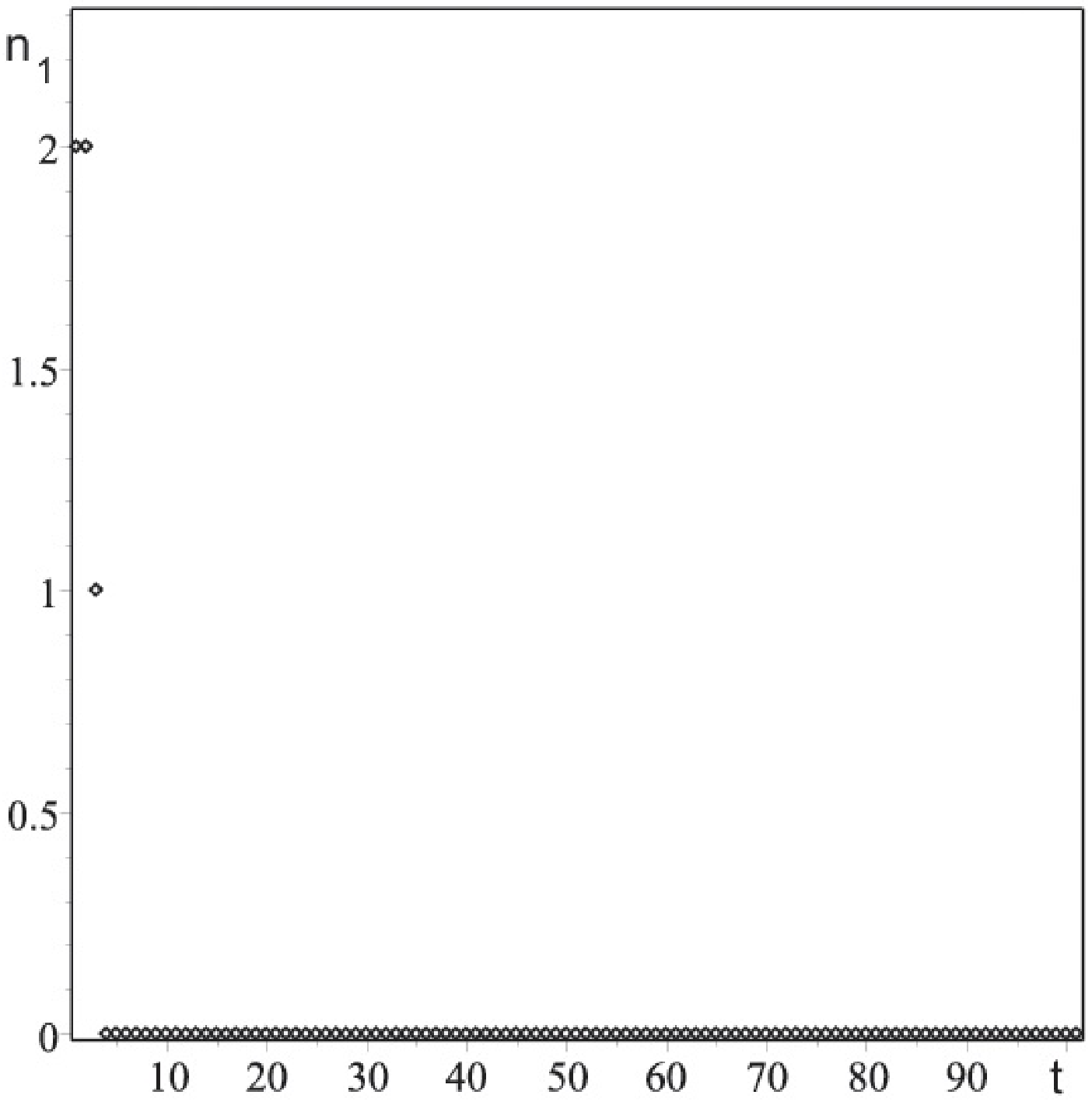}
  \includegraphics[height=4.5 cm]{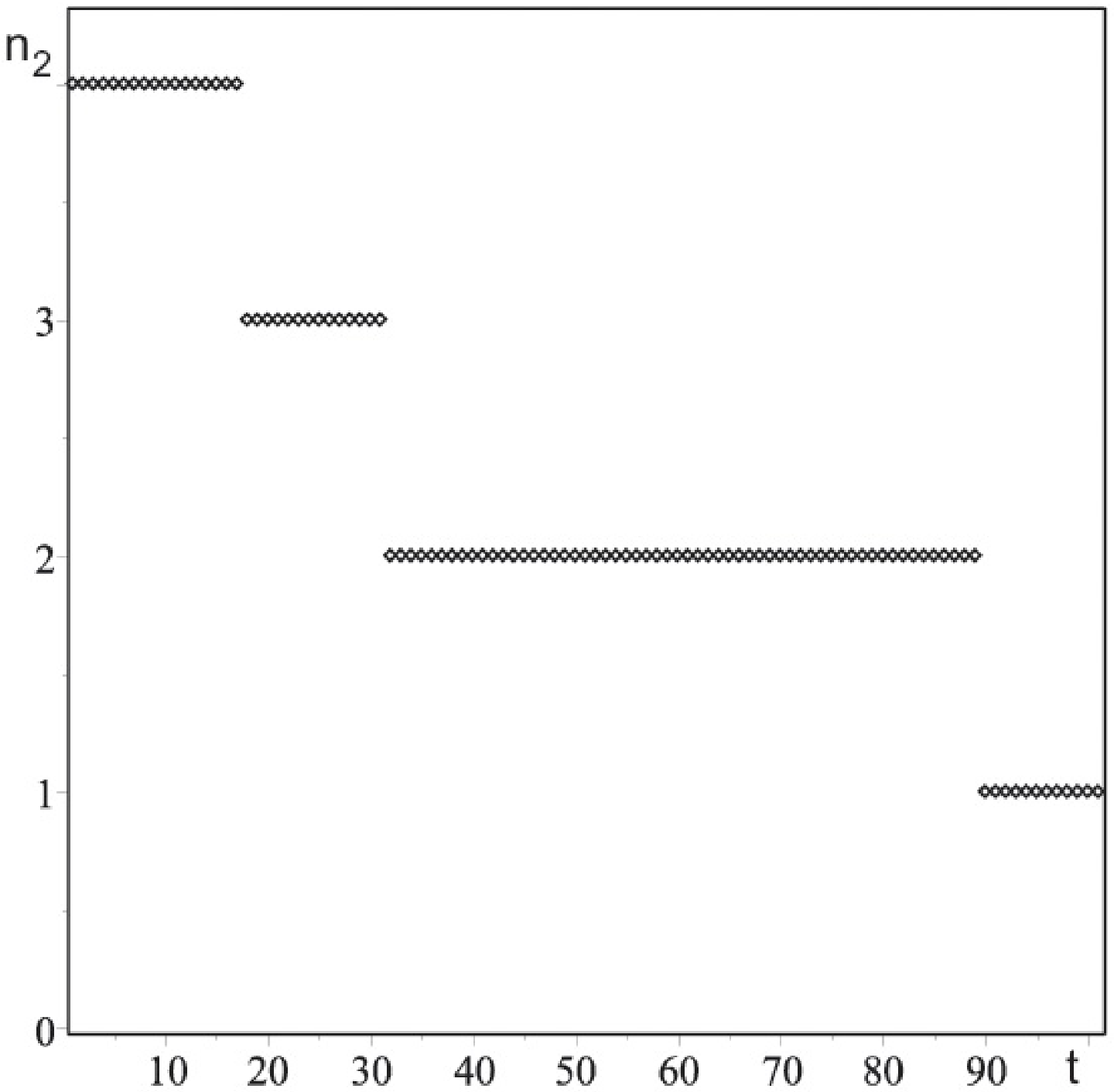}\\
  \includegraphics[height=4.5 cm]{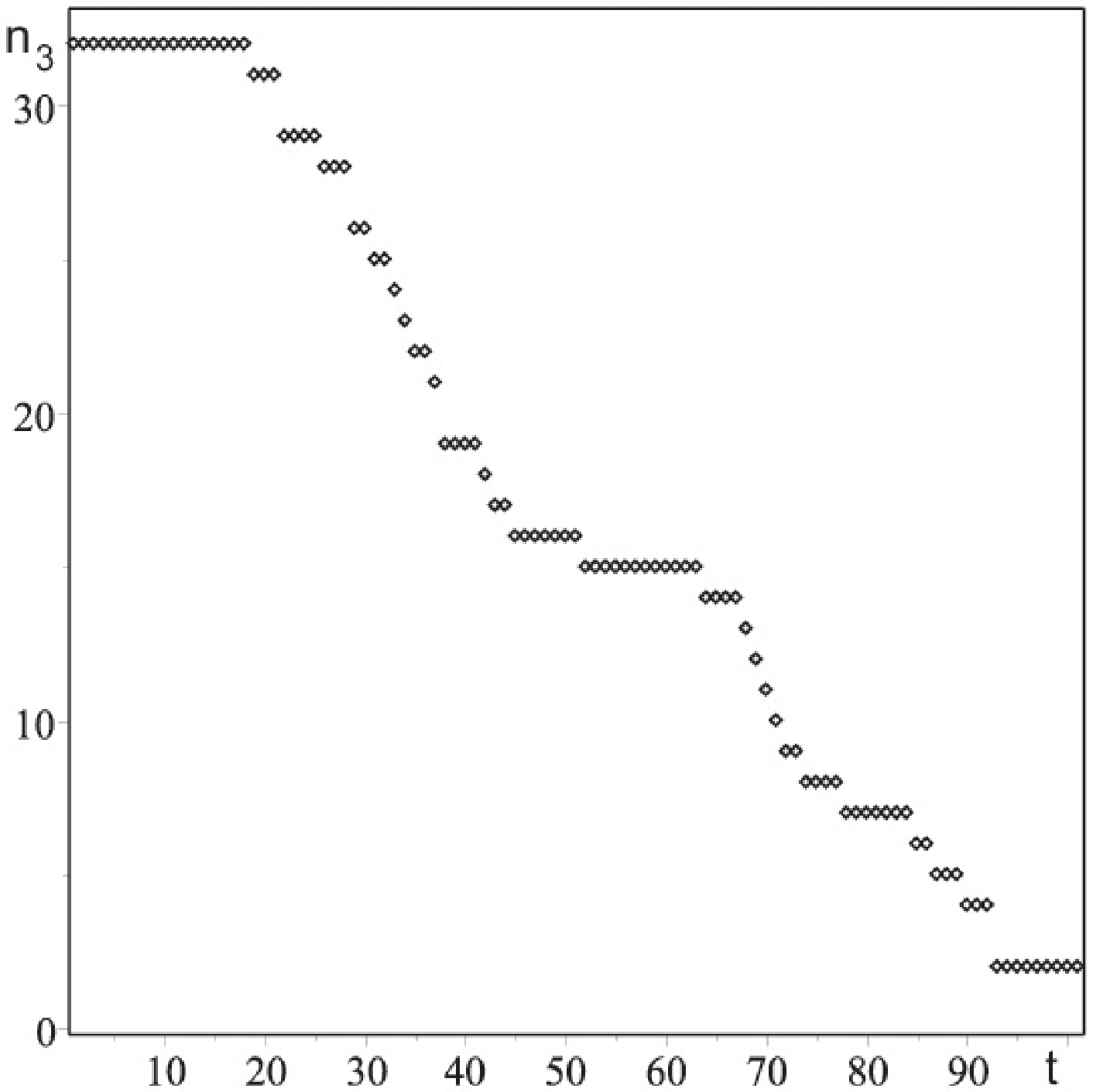}
  \includegraphics[height=4.5 cm]{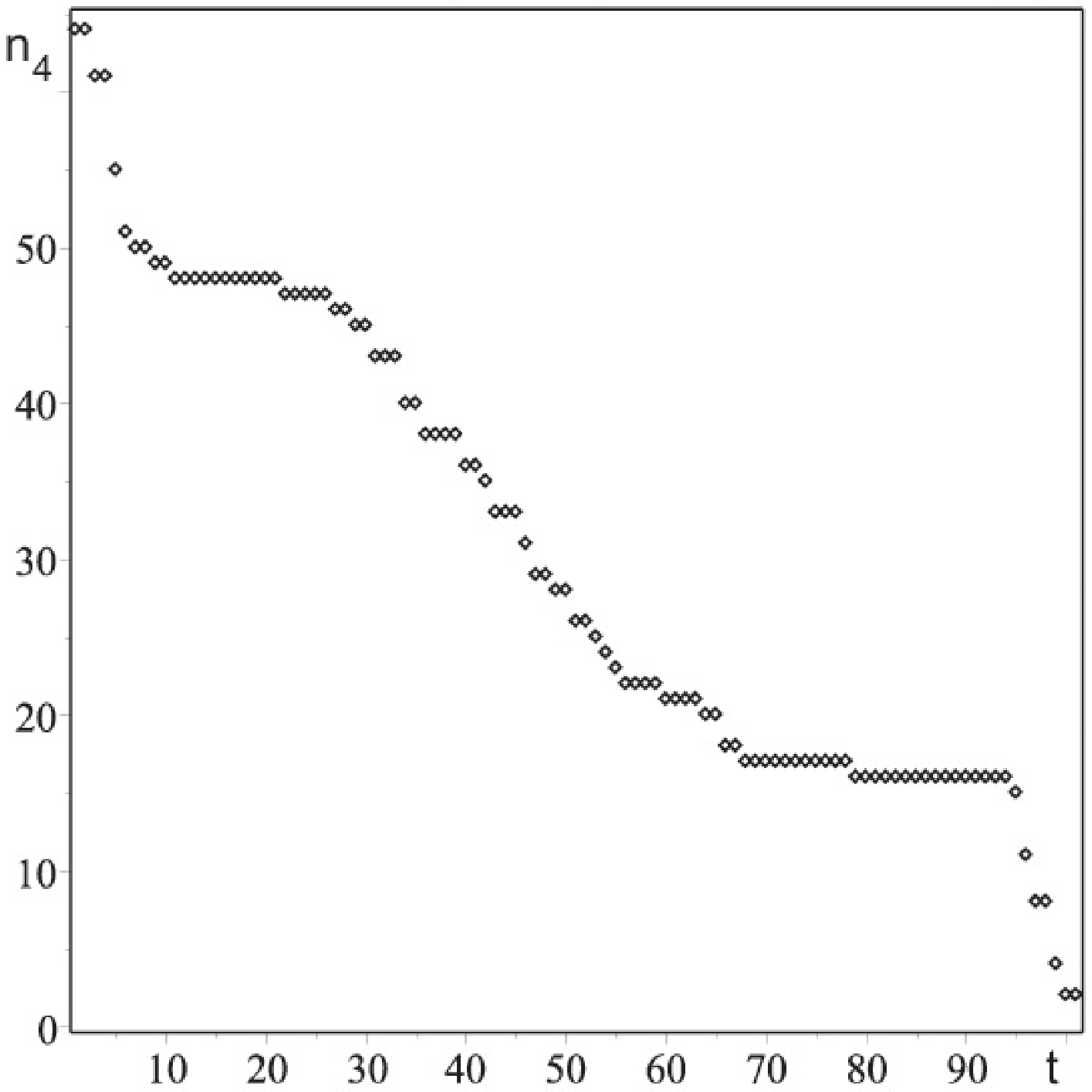}
  \includegraphics[height=4.5 cm]{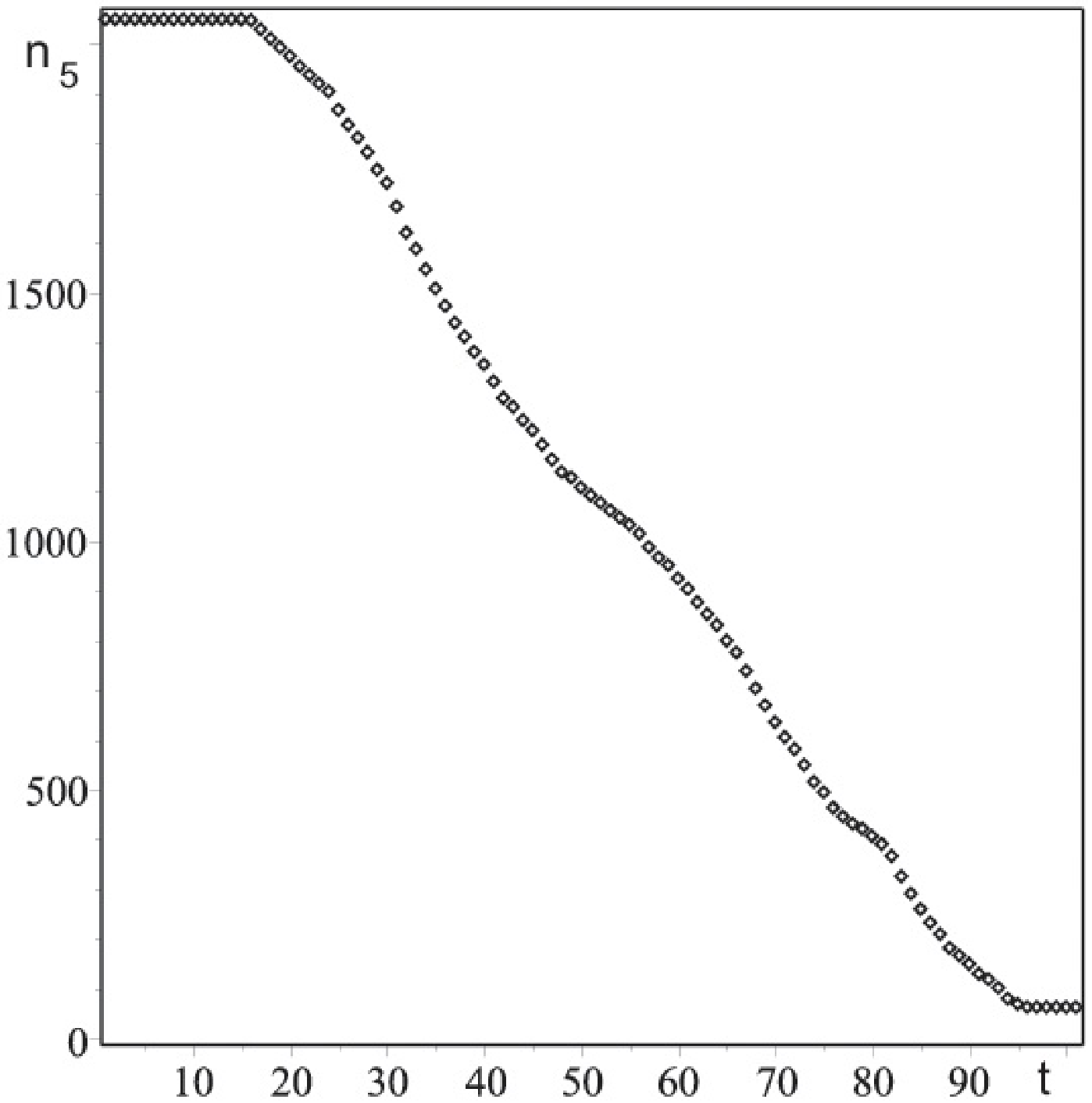}\\
  \includegraphics[height=4.5 cm]{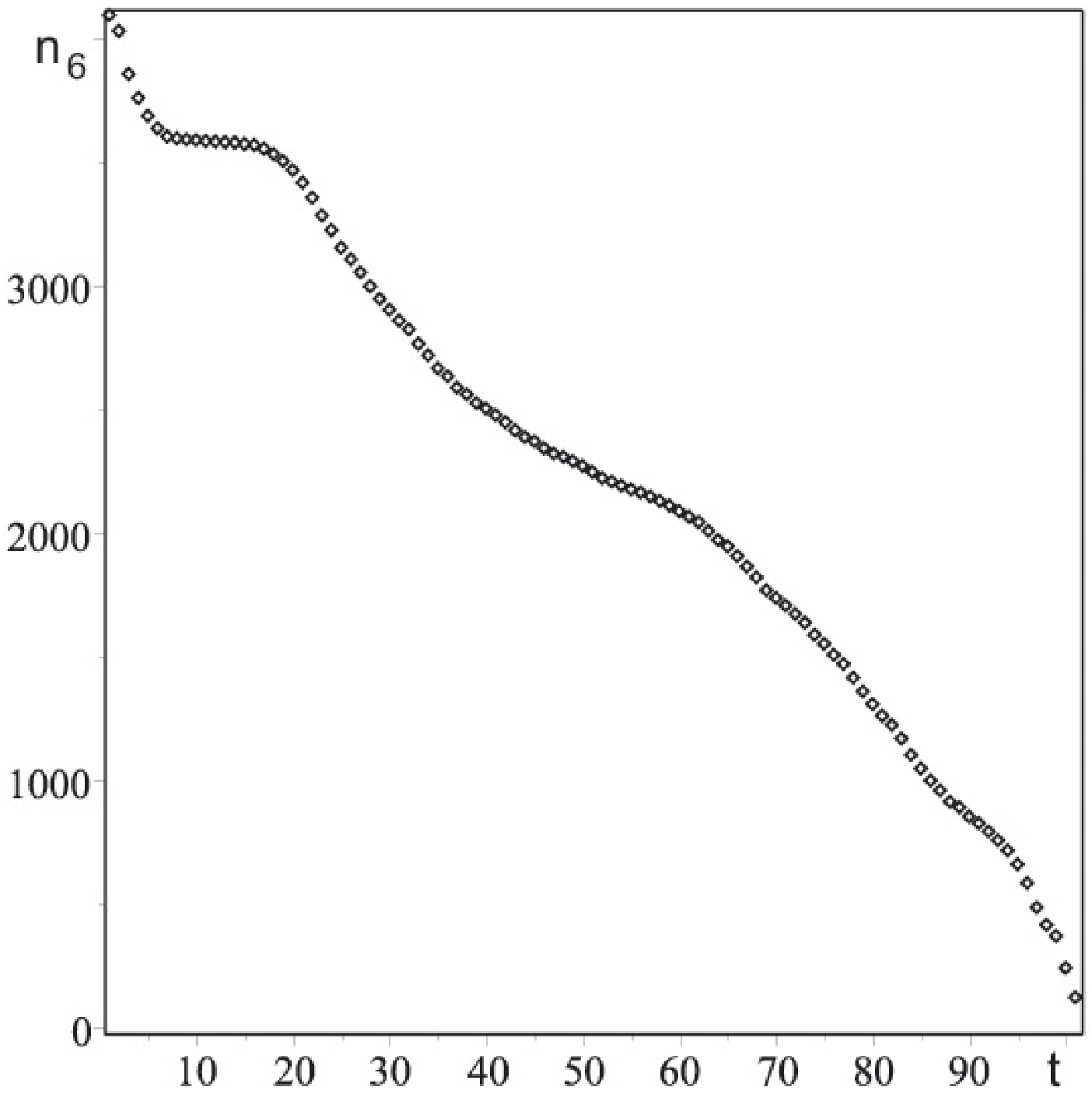}
  \includegraphics[height=4.5 cm]{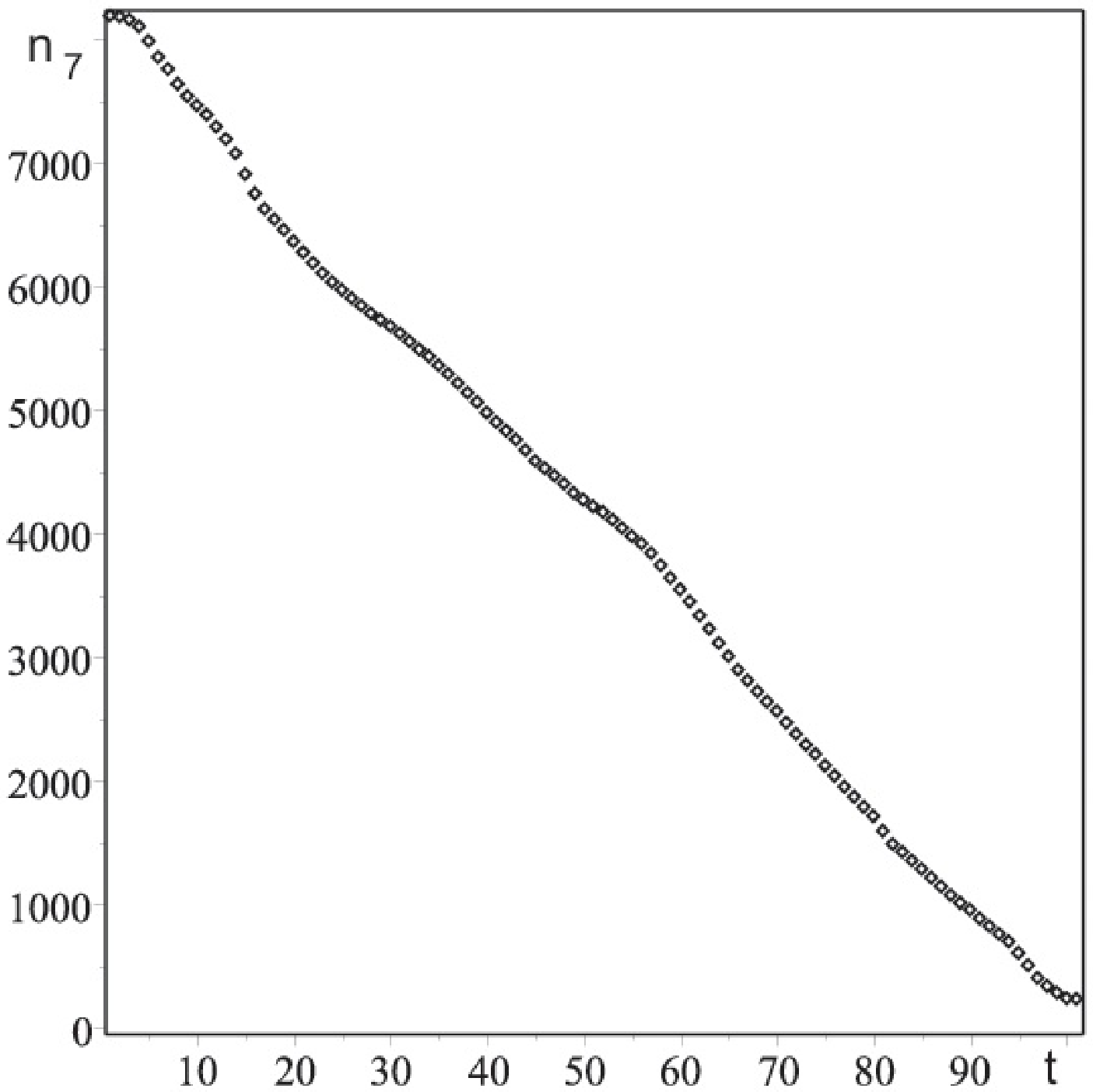}
  \includegraphics[height=4.5 cm]{fig15k.eps}\\
  \caption{Change in the number of strategies of a certain complexity in a world with a depth of memory of 2.}\label{fg15}
\end{figure}

\begin{figure}
  \centering
    \includegraphics[width=5 cm]{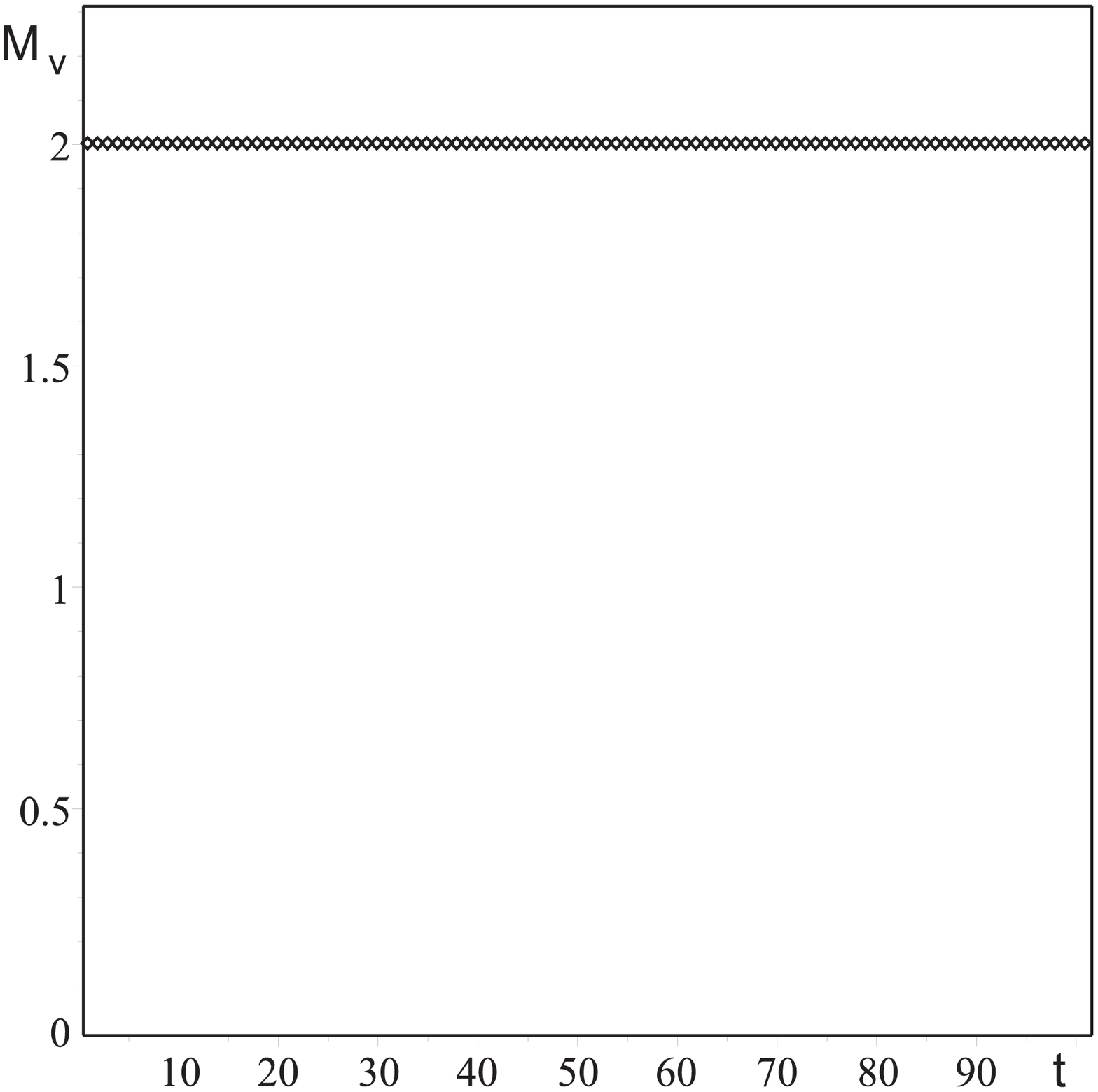}
    \includegraphics[width=5 cm]{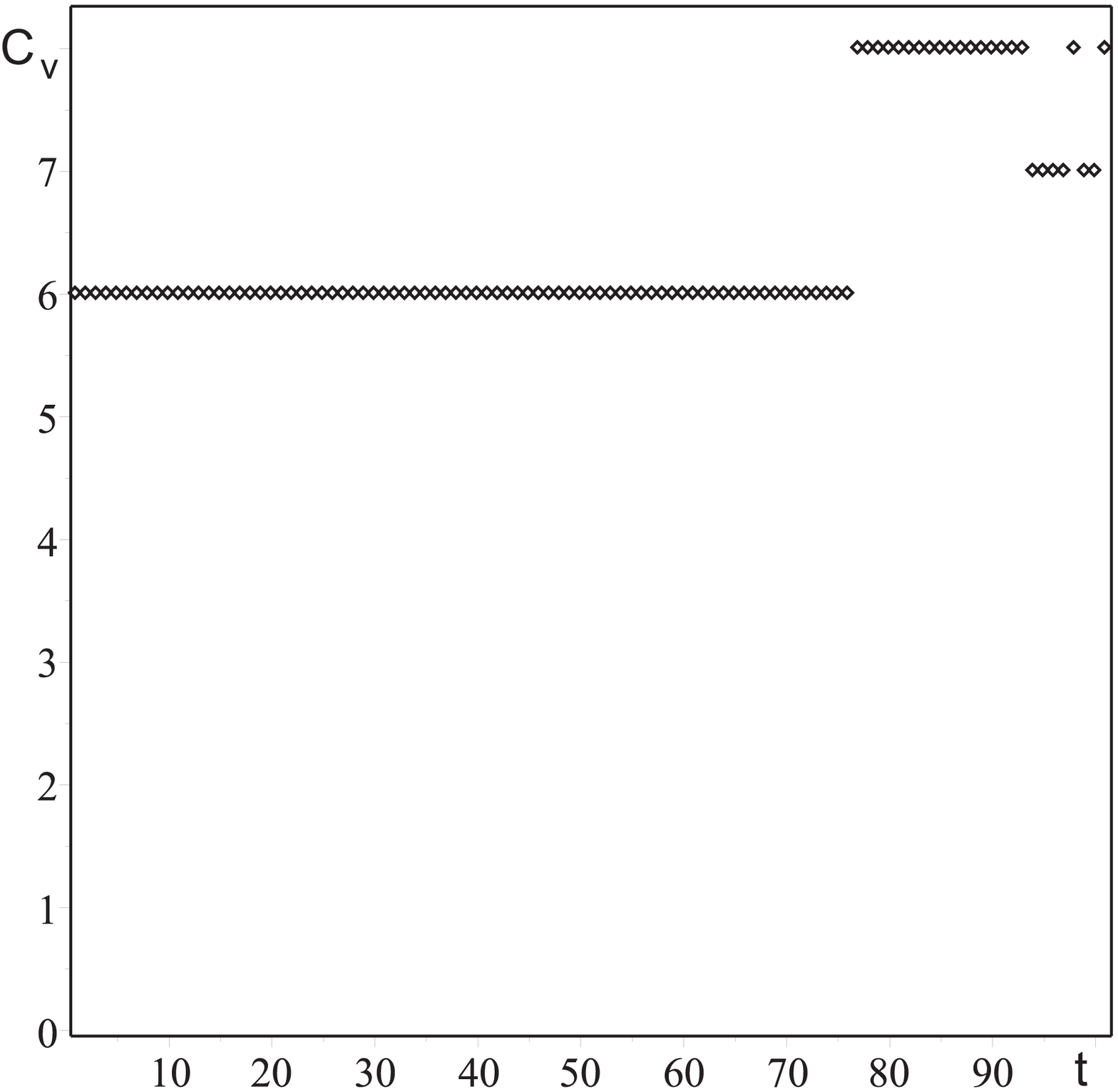}\\
  \caption{Left is the depth of memory of the winning strategy at every stage of evolution. On the right is the complexity of the winner's strategy at every stage of evolution. The data obtained as a result of modeling the evolution of strategies.}\label{fg16}
\end{figure}

The dominance of strategies with a large depth of memory is observed in this world at all stages of evolution (see Fig.\ref{fg16}). Unlike a community without memory and with memory 1, strategies at maximum memory depth win at all stages. Therefore, there is no era of dominance of strategies with zero memory depth. The depth of memory from this point of view is an evolutionarily advantageous property. However, it should be noted that the number of retired strategies from a community with this memory depth (2) exceeds the total number of strategies with a smaller memory depth ($0,1$).

We turn to the behavior of the complexity of strategies over time. Of course, as in the previous cases, we use the collective variables $n_i (t)$. The results of numerical simulation are shown in Fig.\ref{fg15}. These dependencies show that primitive strategies of small complexity disappear from population in the early stages of evolution, not reaching the final stages of the struggle for existence. So, strategies of complexity 1 - at 900 are the first to disappear, the most "decent" strategy disappears even earlier at stage 551. Strategies of the 0th complexity disappear at stage 16500, the most aggressive strategy disappears at stage 16204. Strategies of greater complexity do not completely disappear and form a stationary set of strategies. It is possible to construct the distributions of stationary strategies according to the depth of memory and complexity (see Fig.\ref{fg19}). The hospital has the highest number of the most complex strategies and with maximum memory depth (see Fig.\ref{fg19}).

\begin{figure}
  \centering
    \includegraphics[width=5 cm]{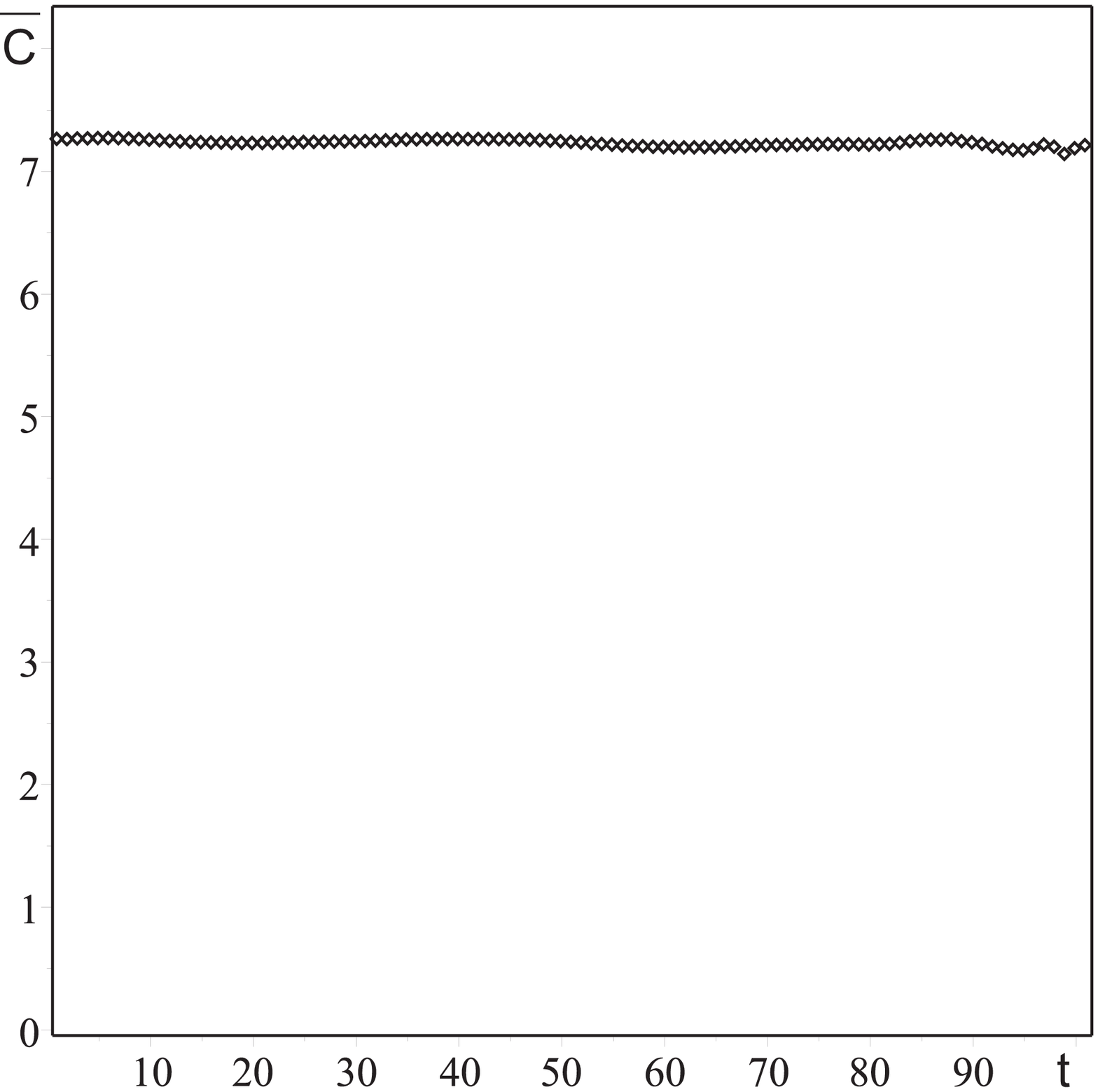}
  \includegraphics[width=5 cm]{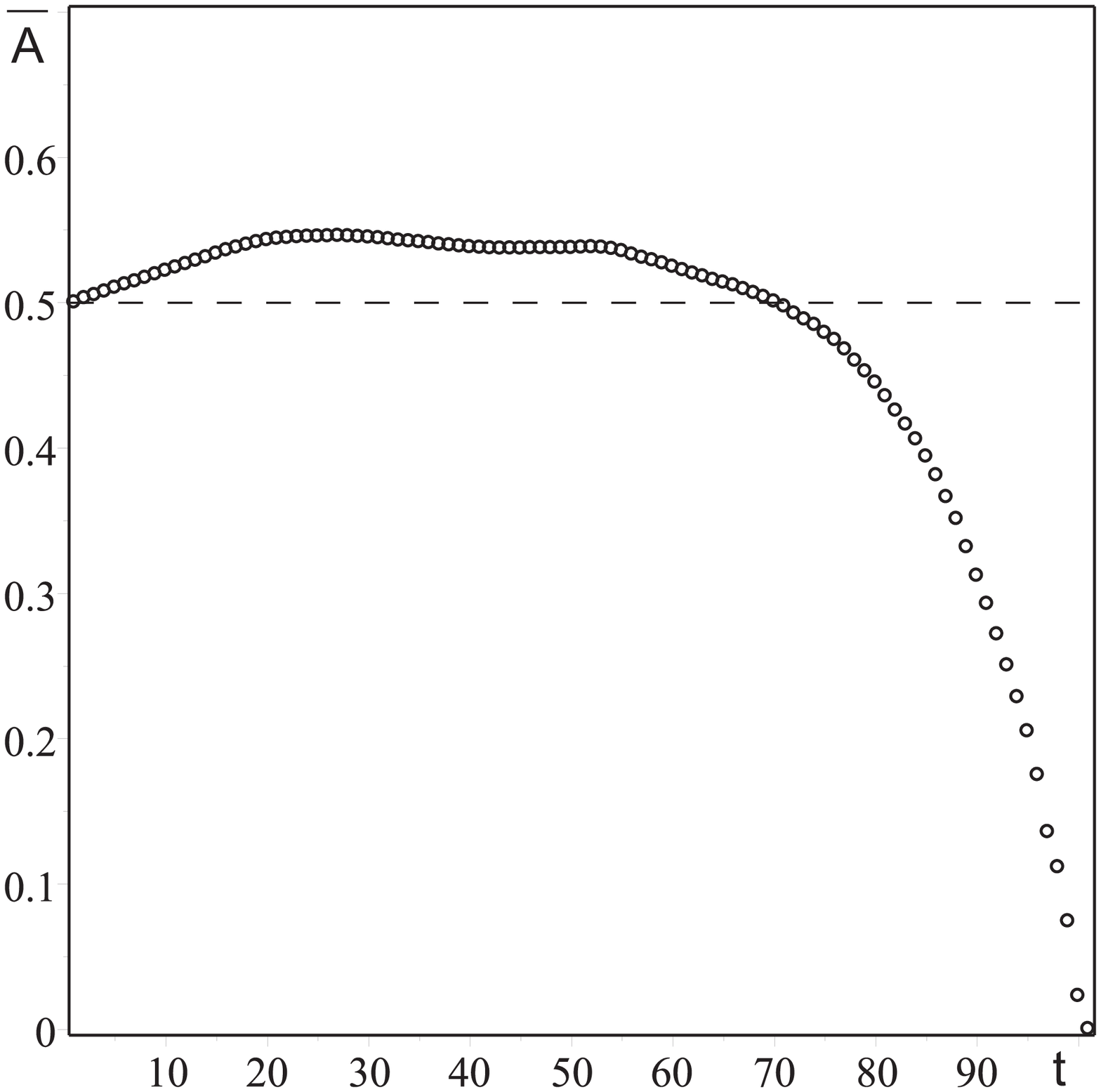}
  \includegraphics[width=5 cm]{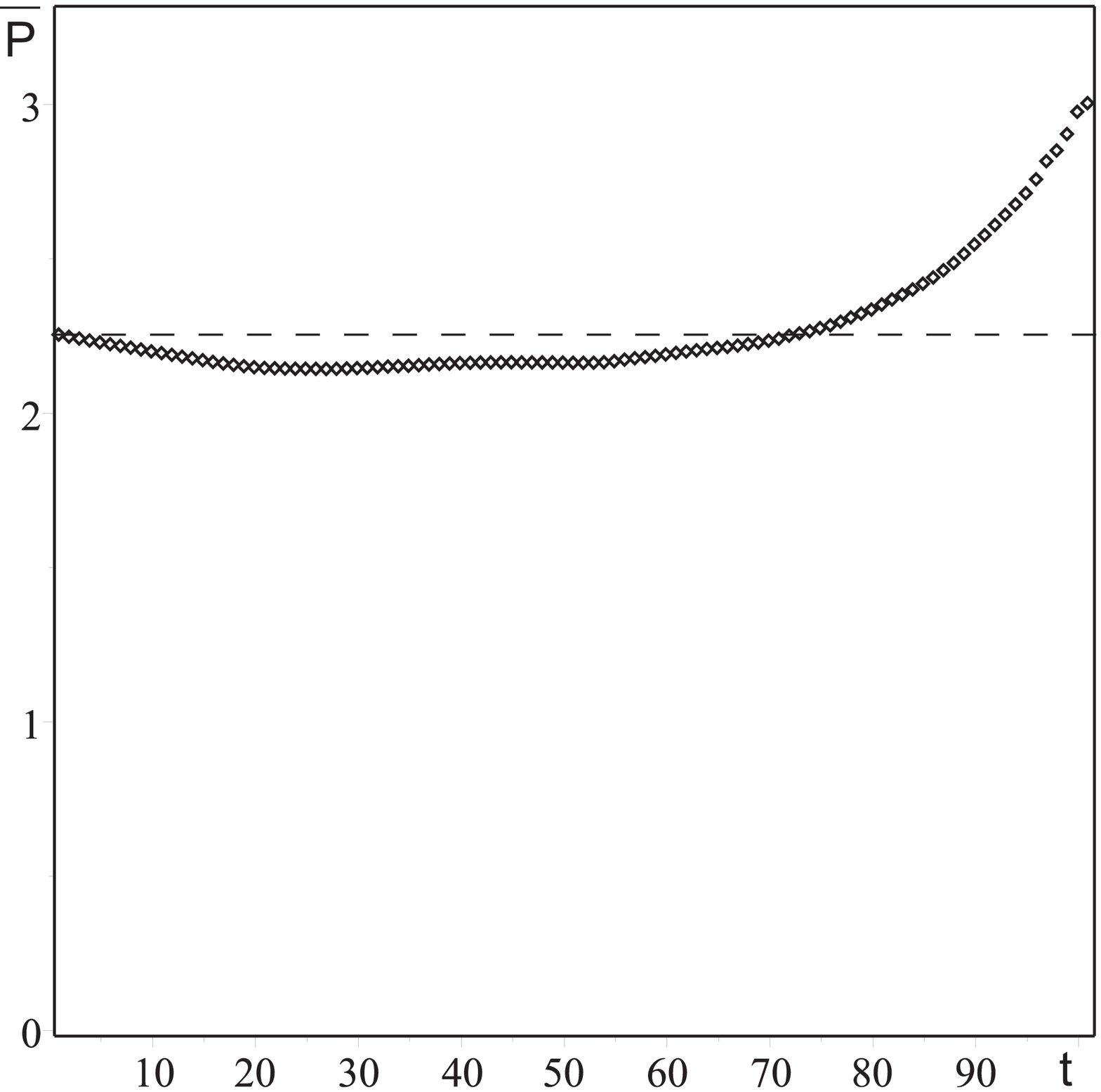}\\
  \caption{Left is the behavior of medium complexity social strategies. In the center is the average aggressiveness of strategies at every stage of evolution. On the right, the number of points on average per turn of one strategy. The dashed line shows the corresponding characteristics of a population in which all strategies with a memory depth of 2 and below are present.}\label{fg17}
\end{figure}

In a world with memory, complexity of strategies is an advantageous property in evolution. It can be said that evolution supports and approves the complexity of strategies. To demonstrate this, one can cite a change in the average complexity of the strategies of the whole population in the process of evolution (see Fig.\ref{fg17}). It can be seen that the average complexity of strategies changes little during evolution and its small oscillations at the final stages of evolution are associated with a decrease in the number of social strategies. In this case, the disappearance of even one strategy affects the average value. It can be assumed that the average value of the complexity of strategies is preserved during the evolution of population and with a greater depth of memory.

In a world with a depth of memory of 2, complex strategies dominate at all stages of evolution. This is clearly seen from Fig.\ref{fg16}, which shows the complexity of the winning strategy at each stage of evolution. It is easy to see that strategies of low complexity are absent among the winners at all stages of evolution. In other words, the period of development of population, when primitive strategies dominate, is absent in this world. Also, the presence of primitive strategies cannot be used to highlight a primitive period. One of them is included in the stationary set of strategies. This is a strategy of zero memory depth and complexity 2.

Figure 17 shows the average aggressiveness of the community at each stage of evolution. A characteristic dependence of average aggressiveness with time is visible. In all worlds with different depths of memory, at first aggressiveness increases, and then decreases to a minimum value. Then it should be considered more natural to single out the primitive period by increasing aggressiveness in population and achieving maximum value. If we consider aggressiveness, then there are two local maximums. One at $27 \cdot 300$, and the second at stage $52 \cdot 300$. Estimating the primitive period by the first maximum gives $27 \%$ of the evolution time. It seems natural to choose a second maximum to highlight a primitive period. Then the primitive period takes $52 \%$ of the time to reach the stationary state. You may notice that the tendency to decrease in the primitive period with increasing memory depth persists. Note that it ends before the disappearance of primitive strategies.

We now turn to a discussion of the evolutionary advantage points obtained on average in one turn by the strategy. The general tendency to decrease points with increasing aggressiveness persists in this world (see Fig.\ref{fg17}). The nature of change has a typical appearance for all worlds. The differences come down to the relative position of the minimum of the obtained points of evolutionary advantage. The correlation between aggressiveness and the number of points per move is also noticeable.
\begin{figure}
  \centering
  \includegraphics[width=6 cm]{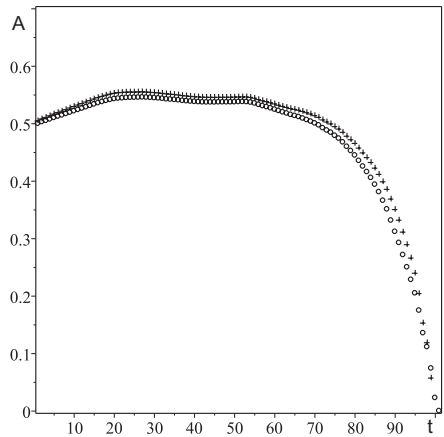}\\
  \caption{Average aggressiveness - squares and aggressiveness constructed according to equation (\ref{q1}) - crosses.}\label{fg18}
\end{figure}

It remains to verify the feasibility of the universal connection (\ref{q1}) between these characteristics. Fig.\ref{fg18} shows the average aggressiveness and aggressiveness, constructed according to the dependence of the number of points per move. The consistency of these dependencies is clearly visible. As memory grows, data matching improves.

It is interesting to note that the selected values of $\bar{P}_{max}=3$, as well as the coefficient $\lambda= 5.3/8$ and $a=0.2$ in the world without memory, did not change for other worlds.

We now turn to a discussion of the stationary state arising in the process of evolution. It is formed by a fairly large number of strategies - 857 strategies

All of them gain the same number of points at the stationary stage and maintain zero aggressiveness in relation to each other. This is a community of "friendly" strategies. The distribution of the number of strategies by memory depth and complexity is shown in Fig.\ref{fg19}. A very small number of strategies have a memory depth less than the maximum. These are only 5 strategies - one with zero memory depth and 4 with memory depth 1. The distribution of strategies by complexity is more meaningful. Starting from complexity 5, the data are well approximated by an exponential dependence. In the hospital, the overwhelming number of strategies has the maximum falsity.
\begin{figure}
  \centering
  \includegraphics[width=5 cm]{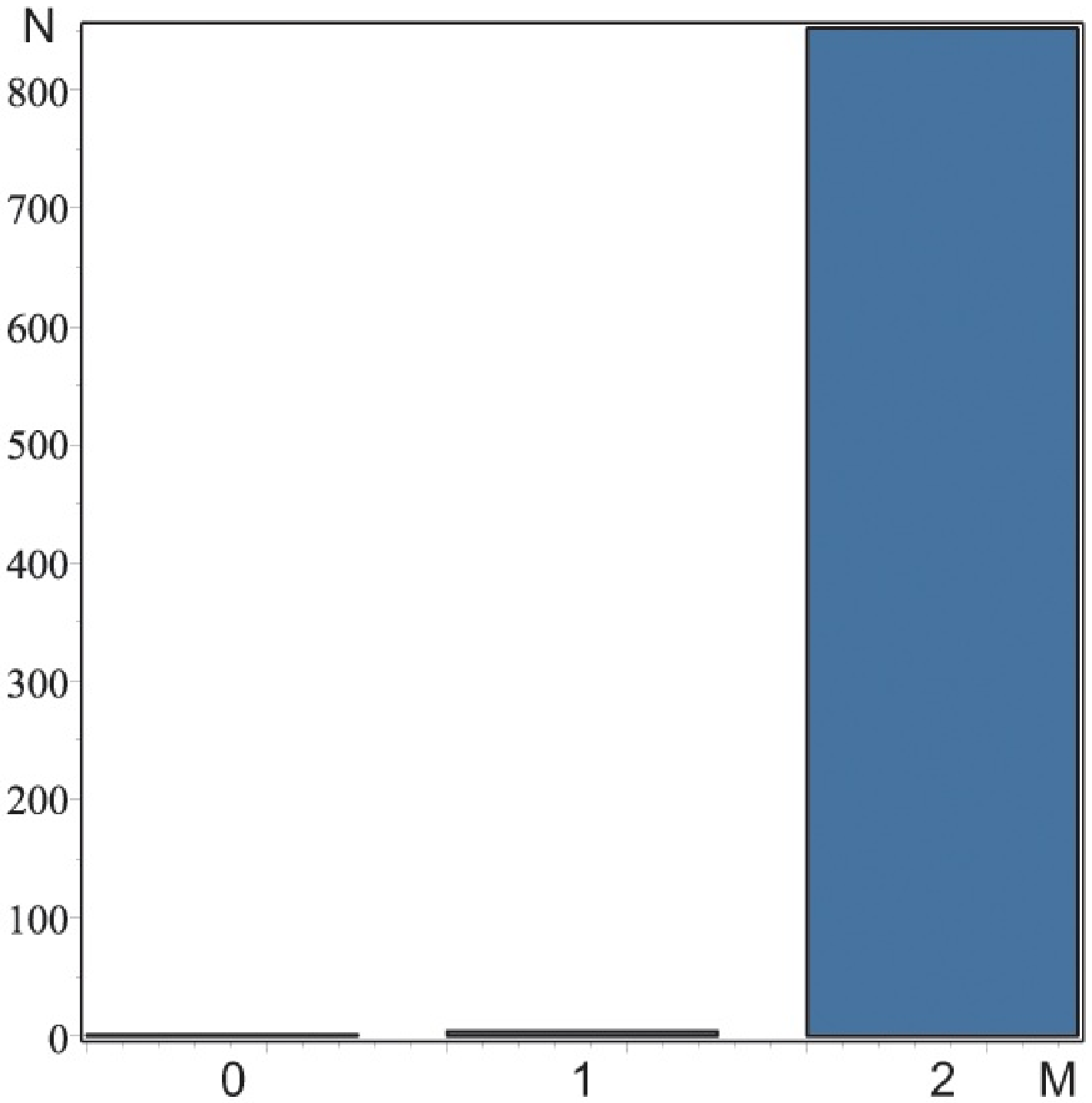}
  \includegraphics[width=5 cm]{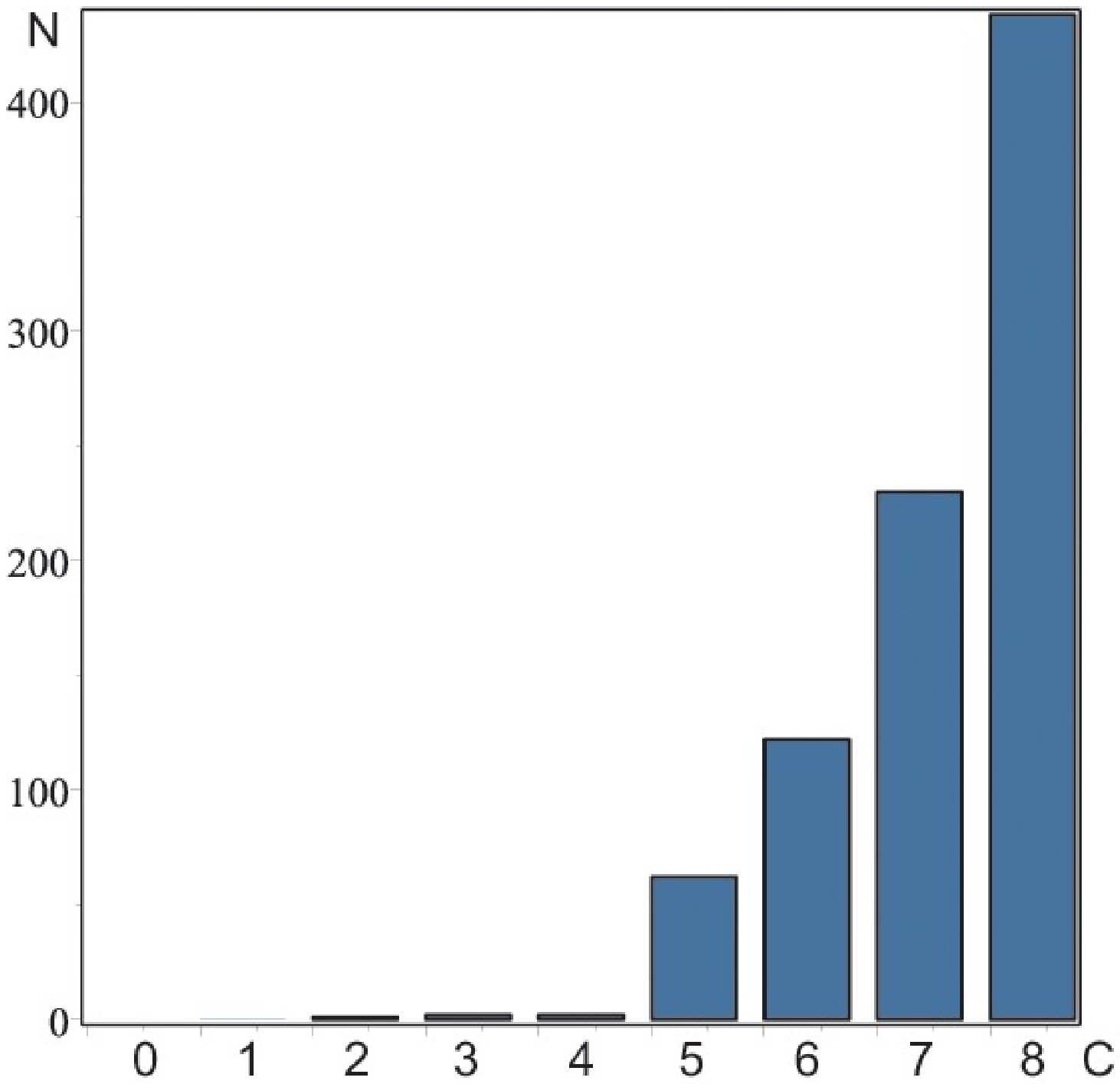}\\
  \caption{The distribution of strategies in the hospital on the left by the depth of memory, on the right by complexity. The vertical axis is the number of strategies.}\label{fg19}
\end{figure}

\section{Conclusion}

First, we note that for the Cauchy problem under consideration, memory and, as a consequence, complexity provide evolutionary advantages. Strategies with little memory and little complexity die out. The average memory and complexity of a population with a fixed depth of strategy memory changes little during evolution and is close to maximum values. Apparently, this is the main reason for the complexity and occurrence of diversity during evolution. In addition, the surviving strategies in the hospital have zero aggressiveness towards each other. In a certain sense, we can say that memory is the universal mechanism for the emergence of cooperation in the community.

In all the considered worlds, one can distinguish a primitive period during which the aggressiveness of strategies in population grows. With increasing depth of memory, the relative duration of this period decreases. So, in the world without memory, the primitive period takes $62.5 \%$ of the time to reach the stationary state, in the world with memory 1 - $37 \div 38 \%$, and in the world with memory 2 - $27 \% \div 52 \%$. Primitive strategies exist in population even after the end of the primitive period. And with a memory depth of 2, one strategy with a memory depth of 0 and complexity 3 reaches the hospital.

The lifetime of the most aggressive strategy in population with the increasing depth of memory is also reduced. So, with a memory depth of 0, it takes $62.5 \%$ of the evolutionary time, in the world with memory 1 - $55 \%$, and in the world with memory 2 - $54 \%$. You can notice a correlation with the length of the primitive period.

In all worlds, the dependence of average aggressiveness on time has a characteristic bell-shaped appearance. The differences are in the position of the maximum and its value. So, with increasing depth of memory, the maximum shifts toward the beginning of evolution, and its value decreases, which makes it difficult to find its position. Therefore, with increasing depth of memory, its width increases. Aggression in the process of evolution after a primitive period decreases and tends to a minimum value. Inpatients form strategies with zero aggressiveness towards each other.
There is a universal connection (\ref{q1}) between the aggressiveness of population and the number of evolutionary advantage points that an average strategy receives per turn. The increase in aggressiveness reduces the amount of payments per move.

\end{document}